\documentclass[fleqn,usenatbib]{mnras}

\usepackage[utf8]{inputenc}

\usepackage{acronym}
\usepackage{ae,aecompl}
\usepackage{booktabs}
\usepackage{dcolumn}
\usepackage{graphicx}
\usepackage{amsmath,amsfonts,amssymb}
\usepackage{mathrsfs}
\usepackage[normalem]{ulem}
\usepackage{epstopdf}

\usepackage[T1]{fontenc}

\let\oldhref\href
\renewcommand{\href}[2]{\oldhref{#1}{\hbox{#2}}}

\title[CCSN neutrinos informed by 3D models]{Core-collapse supernova neutrino emission and detection informed by state-of-the-art three-dimensional numerical models}

\author[H. Nagakura et al.]{
Hiroki Nagakura$^{1}$\thanks{E-mail: hirokin@astro.princeton.edu},
Adam Burrows$^{1}$,
David Vartanyan$^{2}$,
David Radice$^{3,4}$
\\
$^{1}$Department of Astrophysical Sciences, Princeton University, 4 Ivy Lane, Princeton, NJ 08544, USA\\
$^{2}$Astronomy Department and Theoretical Astrophysics Center, University of California, Berkeley, CA 94720, USA\\
$^{3}$Department of Physics, The Pennsylvania State University, University Park, PA 16802, USA\\
$^{4}$Department of Astronomy \& Astrophysics, The Pennsylvania State University, University Park, PA 16802, USA
}

\date{Accepted XXX. Received YYY; in original form ZZZ}

\pubyear{2020}

\begin{document}
\label{firstpage}
\pagerange{\pageref{firstpage}--\pageref{lastpage}}
\maketitle

% Abstract of the paper
\begin{abstract}
Based on our recent three-dimensional core-collapse supernova (CCSN) simulations including both exploding and non-exploding models, we study the detailed neutrino signals in representative terrestrial neutrino observatories, Super-Kamiokande (Hyper-Kamiokande), DUNE, JUNO, and IceCube. We find that the physical origin of difference in the neutrino signals between 1D and 3D is mainly proto-neutron-star (PNS) convection. We study the temporal and angular variations of the neutrino signals and discuss the detectability of the time variations driven by the spiral Standing Accretion Shock Instability (spiral SASI) when it emerges for non-exploding models. In addition, we determine that there can be a large angular asymmetry in the event rate ($\gtrsim 50 \%$), but that the time-integrated signal has a relatively modest asymmetry ($\lesssim 20 \%$). Both features are associated with the lepton-number emission self-sustained asymmetry (LESA) and the spiral SASI. Moreover, our analysis suggests that there is an interesting correlation between the total neutrino energy (TONE) and the cumulative number of neutrino events in each detector, a correlation that can facilitate data analyses of real observations. We demonstrate the retrieval of neutrino energy spectra for all flavors of neutrino by applying a novel spectrum reconstruction technique to the data from multiple detectors. We find that this new method is capable of estimating the TONE within the error of $\sim$20\% if the distance to the CCSN is $\lesssim 6$ kpc.
\end{abstract}

% Select between one and six entries from the list of approved keywords.
% Don't make up new ones.

% which are only accessible by multi-dimensional models

\section{Introduction}\label{sec:intro} 
A core-collapse supernova (CCSN) arises from a catastrophic death of a massive star ($M \gtrsim 8 M_{\sun}$). During the development of the explosion and the cooling of the proto-neutron star (PNS), a total energy of $\sim$3 $\times 10^{53} {\rm erg}$ is radiated. Indeed, a neutrino burst associated with a CCSN was directly detected from SN 1987A by the terrestrial neutrino detectors Kamiokande \citep{1987PhRvL..58.1490H} and the IMB (Irvine–Michigan–Brookhaven) \citep{1987PhRvL..58.1494B}. They detected a total of $19$ events with neutrino energies ranging from $\sim 5$ MeV to $\sim 40$ MeV and the event lasted $\sim 10$ seconds; this is consistent with our understanding of the dynamics of CCSN, albeit very crudely. On the other hand, due to low number of events and poor flavor sensitivity, these observations did not provide enough information to constrain the explosion mechanism . This is a major opportunity for the future.

A large number of neutrino detectors with sensitivities to multiple neutrino flavors should be available when the next galactic CCSN happens. Super-Kamiokande (SK), one of the currently operating water-Cherenkov neutrino detectors, is capable of detecting $\sim 10,000$ neutrinos from a CCSN at the distance of $10$ kpc \citep{2016APh....81...39A}; Hyper-Kamiokande (HK) is essentially a scaled-up version of SK by a factor of several in volume \citep{2018arXiv180504163H} and the project has recently been officially approved. The deep underground neutrino experiment (DUNE) is a liquid-Argon detector which will have a unique sensitivity to electron-type neutrinos ($\nu_e$) \citep{2016arXiv160105471A,2016arXiv160807853A}. The Jiangmen Underground Neutrino Observatory (JUNO), one of the upcoming liquid-scintillator detectors, is designed mainly to determine the neutrino mass hierarchy by a precise measurement of reactor antineutrino spectra, but this capability is also useful for detecting CCSN neutrinos \citep{2016JPhG...43c0401A}. Due to its large effective mass, IceCube will detect $\sim 100$ times the number of neutrino events that SK will \citep{2011A&A...535A.109A}\footnote{We refer readers to e.g., \citet{2012ARNPS..62...81S} and references therein for other available methodologies capable of addressing the detection of CCSN neutrinos.}. Taking advantage of these facilities, the time structures and flavor-dependent features in neutrino signals are available for observation, and such data can help us understand not only CCSN dynamics, but neutrino flavor conversion in CCSN cores.

CCSN dynamics is governed by the complex interplay between microphysics and macrophysics, which engenders rich structures in its neutrino signal. With a future direct detection of supernova neutrinos, the complex features therein encoded will be best studied by comparing with theoretical predictions of the signals obtained by the sophisticated numerical simulations. There have been many such studies \citep[see, e.g.,][]{2012ARNPS..62...81S,2013ApJ...762..126O,2019ApJ...881..139S,2019PhRvD..99l3009L,2019arXiv191203328W}, and groups have used either their own CCSN simulations or publicly available models \citep{2010PhRvL.104y1101H,2013ApJS..205....2N}. However, these neutrino data generally derive from spherically-symmetric CCSN simulations. More importantly, such 1D studies generally employ artificial means to explode the model and to determine when to explode it (see \citet{2010PhRvL.104y1101H} for an exception). Such an approach is unavoidable, since shocks do not generally revive in state-of-the-art 1D simulations. It is, therefore, preferable when developing theoretical predictions of neutrino signals to conduct multi-dimensional simulations for which explosion is a common outcome.

In this paper, we present the first systematic study of neutrino signals based on our 3D CCSN simulations. Although previously there have been a few explorations of the neutrino emissions from 3D models \citep[see, e.g.,][]{2018MNRAS.475L..91T,2019MNRAS.489.2227V,2019ApJ...881...36G,2019PhRvD.100f3018W,2019arXiv191012971W,2019ARNPS..6901918M}, little has been studied of the differences between 3D and 1D models in a detector.  In this paper, we develop a new analysis pipeline by employing and extending the SNOwGLoBES detector software\footnote{See \url{https://webhome.phy.duke.edu/~schol/snowglobes/} for more details} and we discuss the unique characteristics of 3D models, in particular their expected neutrino event rate, energy spectrum, angular dependences and time variation. We use the F{\sc{ornax}} code \citep{2019ApJS..241....7S} in our 3D CCSN simulations. F{\sc{ornax}} incorporates the important neutrino-matter interactions and multi-group neutrino transport, taking into account the fluid-velocity dependence and general-relativistic redshifts. The relatively speedy F{\sc{ornax}} code has allowed us to conduct many CCSN simulations across a wide range of massive-star progenitors, with both exploding and non-exploding outcomes. Although these CCSN models are still provisional and need further improvements, the essential characteristics of the progenitor dependence may have been captured adequately. All the numerical neutrino data we use in this paper are publicly available\footnote{from the link, \url{https://www.astro.princeton.edu/~burrows/}}.

This paper is organized as follows. In Sec.~\ref{sec:methodsandmodels}, we briefly summarize our numerical CCSNe models. We then provide basic information for calculating neutrino signals in the detector for normal-hierarchy and inverted-hierarchy neutrino oscillation models. All results of our analysis are presented in Sec.~\ref{sec:result}. Finally, in Sec.~\ref{sec:sumconc} we wrap up with a summary and conclusions.

\section{Methods and models}\label{sec:methodsandmodels} 

\subsection{3D CCSN Models}\label{subsec:3DCCSNmodel} 

\begin{figure}
  \begin{minipage}{1.0\hsize}
        \includegraphics[width=\columnwidth]{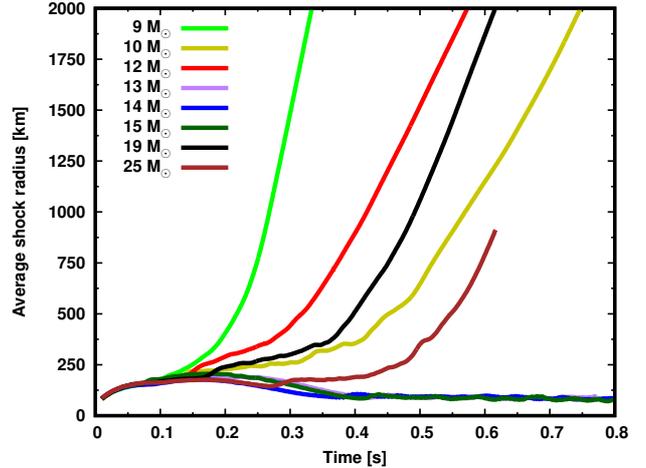}
    \caption{The time evolution of the mean shock radius for 3D models upon which we focus in this paper. Color distinguishes the ZAMS mass of the progenitor.}
    \label{graph_timetrajectories_shockradii_neutsig}
  \end{minipage}
\end{figure}

\begin{figure}
  \begin{minipage}{1.0\hsize}
        \includegraphics[width=\columnwidth]{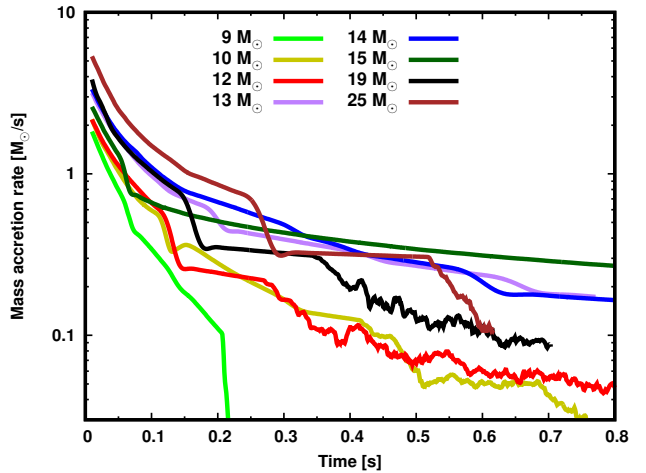}
    \caption{The time evolution of the mass accretion rate measured at 500 km.}
    \label{graph_Massaccretion_neutsig}
  \end{minipage}
\end{figure}

\begin{figure*}
  %\begin{minipage}{0.9\hsize}
  \begin{minipage}{0.8\hsize}
        \includegraphics[width=\columnwidth]{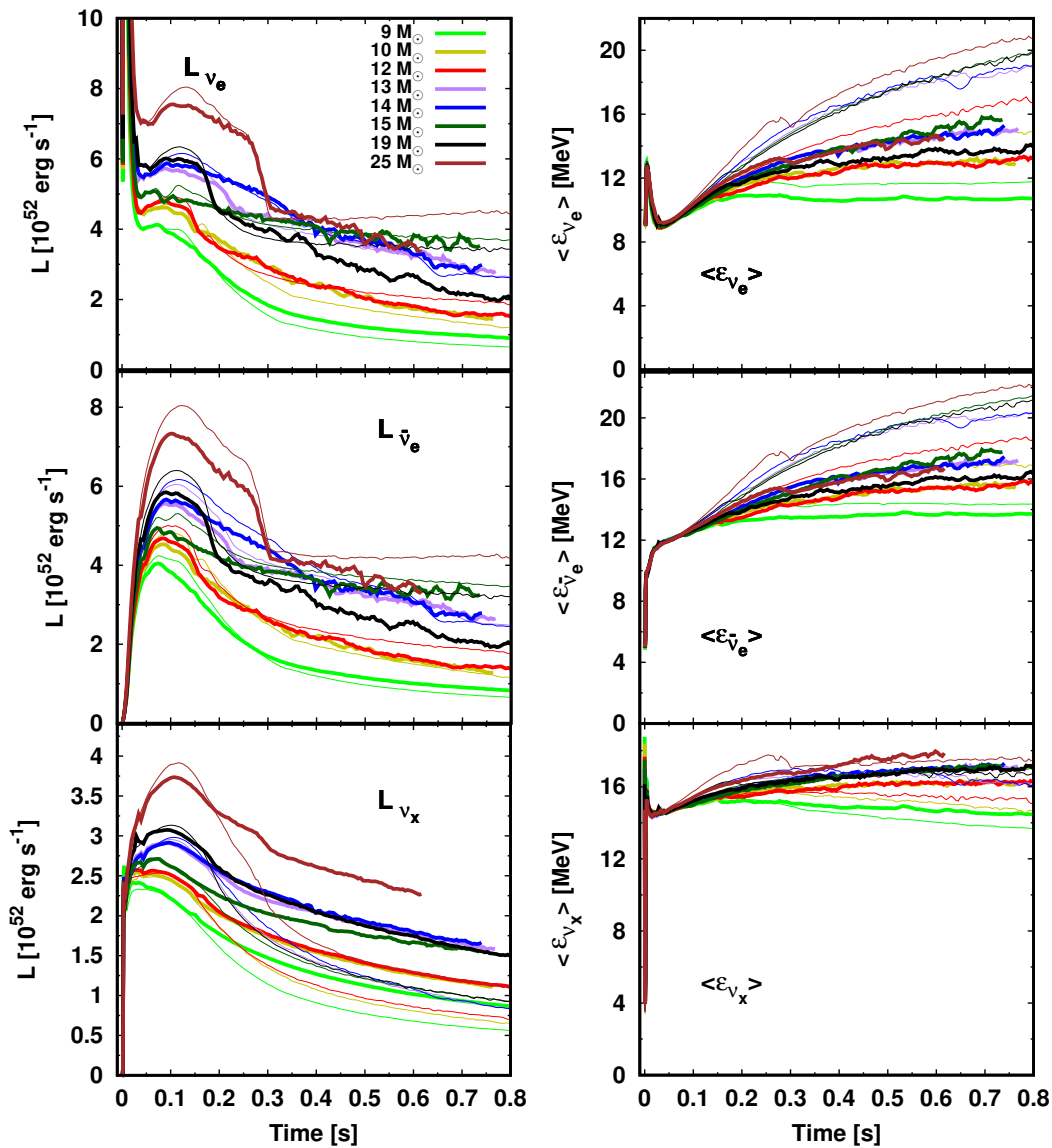}
    \caption{The energy luminosity (left) and average neutrino energy (right) as a function of time for each
species of neutrino for all 3D models in this study. These are measured in the laboratory frame at $250$ km. For comparison, we display 1D counterparts as thin lines. }
    \label{graph_neutrino_lumi_aveene_neutsig}
  \end{minipage}
\end{figure*}

All technical details of the F{\sc{ornax}} code that we employed to generate our 3D CCSN models have been published in a series of papers \citep{2016ApJ...831...81S,2017ApJ...850...43R,2018MNRAS.477.3091V,2019ApJS..241....7S}.  We refer readers to \citet{2019MNRAS.485.3153B,2020MNRAS.491.2715B} for a detailed discussion of these 3D models, \citet{2019MNRAS.489.2227V} for an exploration of the temporal an angular variations of the neutrino signals, and \citet{2020MNRAS.492.5764N} for an analysis of PNS convection.

At the onset of collapse, our CCSN simulations are performed in spherical symmetry by taking a matter profile computed by stellar evolution models. In this study, we include the results for 9-, 10-, 12-, 13-, 14-, 15-, 19- and 25-$M_{\sun}$ models (8 models in total), enough to blanket the overall progenitor dependence\footnote{All progenitor models employed in this paper are non-rotating models.}. The initial 1D models were calculated by  \citet{2016ApJ...821...38S}, except for the 25-$M_{\sun}$ progenitor which was calculated by \citet{2018ApJ...860...93S}. Once the simulation reached $10$ ms after core bounce, we mapped both matter and neutrino radiation profiles to 3D and imposed non-radial perturbations in the fluid velocity following the prescription in \citet{2015MNRAS.448.2141M}. We employed a spherical coordinate, dendritic mesh with $678 \times 128 \times 256$ ($r \times \theta \times \phi$) grid points covering $0 \leq r \leq 20,000$ km. The radial grid is logarithmically stretched outside the inner $\sim 70$ km.

The neutrino transport module in F{\sc{ornax}} solves the energy-dependent two moment equations for three neutrino species: electron-type neutrinos $\nu_e$, electron-type anti-neutrinos $\bar{\nu}_e$, and all the other heavy neutrinos bundled into what we call ``$\nu_x$." The fluid-velocity dependence is included up to $\mathcal{O}(v/c)$ and the effect of general relativity is approximate included using the scheme in \citet{2002A&A...396..361R}. We use 12 energy groups which are logarithmically distributed from $1$ MeV to $300$ MeV for $\nu_e$ and $1$ MeV to $100$ MeV for the other species (We note that we display the result of 1D simulation for comparison, in which we employ 20 energy groups for them. The energy-resolution dependence in 1D models are also displayed in Appendix~\ref{sec:resodepe}.). The transport equation is solved with up-to-date neutrino matter interactions detailed in \citet{2006NuPhA.777..356B}, adding recent improvements, e.g., many body corrections \citep{2017PhRvC..95b5801H}. Other modules such as hydrodynamics (and gravity) are solved simultaneously with neutrino transport and energy, momentum, and lepton number coupling in fully and self-consistently included each (see \citet{2019ApJS..241....7S} for more details).

We witnessed successful explosions for the 9-, 10-, 12-, 19- and 25-$M_{\sun}$ models (i.e., the 13-, 14-, 15-$M_{\sun}$ models did not explode).  These outcomes are collectively demonstrated via the time trajectories of the mean shock radii displayed in Fig.~\ref{graph_timetrajectories_shockradii_neutsig}. As is clearly seen, the explodability and the vigor of shock expansion are not monotonic functions of ZAMS mass. It is important to note that the 9-$M_{\sun}$ model is launched earliest ($\lesssim 150$ ms after bounce) and that hydrodynamic instabilities did not have much time to grow to full vigor. Indeed, its explosion geometry is nearly spherical and the multi-dimensional effects for this model are not pronounced. However, we found that, despite the near sphericity of the blast, the neutrino emissions of the 9-$M_{\sun}$ model are asymmetric \citep{2019MNRAS.489.2227V}. The other models have different CCSN dynamics.  For them, neutrino-driven convection is well-developed prior to explosion (if the model explodes). The associated increase in the turbulent pressure and neutrino heating rate can help facilitate shock expansion.

We point out some important features that can be realized only in 3D models. Shock revival is not achieved in 1D models; hence, previous studies have needed to use artificial prescriptions when studying the neutrino signals of CCSN models. For instance, in \citet{2013ApJS..205....2N} shock revival is assumed to take place at a certain time, at which time the simulations are mapped to those of PNS cooling. In this prescription, mass accretion onto the PNS is artificially shut off. As shown in Fig.~\ref{graph_Massaccretion_neutsig}, mass accretion continues after shock revival, supplying mass and power to the PNS and influencing the neutrino emissions quantitatively. Another crucial deficiency of 1D models is that PNS convection is suppressed. In our previous paper \citep{2020MNRAS.492.5764N}, we confirmed that PNS convection commonly occurs for all stellar collapses within the mass range explored, including not only exploding models, but also their failed counterparts. The latter would go through the PNS phase, but subsequently collapse to a black hole. We find that PNS convection substantially affects both the neutrino luminosity and average neutrino energy (see Fig.~\ref{graph_neutrino_lumi_aveene_neutsig}). This is mainly attributed to the fact that the convection changes the quasi-hydrostatic structure of the core and expands the neutrinospheres. These adjustments change the thermodynamic state and thereby affect neutrino emissions (see Sec. 3.4 in \citet{2020MNRAS.492.5764N} for more details). Although convection can be treated approximately in 1D simulations with a mixing length theory \citep[see, e.g.,][]{2012PhRvL.108f1103R}, such an approach can provide only crude qualitative trends and may not be very accurate. It is also important to note that the lepton-number emission self-sustained asymmetry \citep[LESA;][]{2014ApJ...792...96T} is also observed in 3D models and that this is tied to the derived directional variations of the theoretical neutrino signals. Furthermore, we have found that the spiral Standing Accretion Shock Instability \citep[spiral SASI;][]{2007ApJ...656..366B} appears in non-exploding models and that this induces temporal modulation of the neutrino signal and asymmetric neutrino emission. As such, our 3D models capture key multi-dimensional features unavailable using 1D models . We delve into the observational consequences in Sec.~\ref{sec:result}.

\subsection{Detectors}\label{subsec:detectors} 
For this study, we employ the SNOwGLoBES detector software to estimate neutrino event rates and energy spectra in each detector: SK (HK), DUNE, JUNO, and IceCube. This software is publicly available, and it has been widely used by the cognizant community. In this section, we briefly summarize some essential properties of each detector installed in SNOwGLoBES and refer readers to \citet{2012ARNPS..62...81S,2018MNRAS.480.4710S} for a more detailed discussion.

SK is a currently-operating water Cherenkov neutrino detector, and we assume that its volume is 32.5 ktons for the purpose of CCSN neutrino analysis. It is sensitive primarily to $\bar{\nu}_e$s through the inverse-beta-decay reaction on protons (IBD-p):
\begin{eqnarray}
\bar{\nu}_e  + p \rightarrow e^{+} + n\, .
\label{eq:ibdp}
\end{eqnarray}
Elastic scattering on electrons (eES),
\begin{eqnarray}
\nu_i  + e^{-} \rightarrow \nu_i  + e^{-}\, ,
\label{eq:esce}
\end{eqnarray}
is sensitive to all neutrino species, and can be employed to retrieve the neutrino energy spectra of all flavors of neutrinos emitted at the CCSN source (see Sec.~\ref{subsec:retrieving} for more details). Although we employ only IBD-p and eES reaction channels in this study, there are other useful reaction channels for the study of CCSN neutrinos. Another IBD reaction, this with oxygen:
\begin{eqnarray}
\bar{\nu}_e  + {^{16}{\rm O}}  \rightarrow e^{+} + {^{16}{\rm N}}\, ,
\label{eq:ibdObar}
\end{eqnarray}
is capable of detect tagging $\bar{\nu}_e$s, and the charged-current reaction with oxygen is sensitive to $\nu_e$s through the reaction:
\begin{eqnarray}
\nu_e  + {^{16}{\rm O}}  \rightarrow e^{-} + {^{16}{\rm F}}\, .
\label{eq:ibdO}
\end{eqnarray}
We note that all neutrino species experience inelastic-scattering with oxygen via the neutral-current reaction: 
\begin{eqnarray}
\nu_i  + {^{16}{\rm O}}  \rightarrow \nu_i  + {^{16}{\rm O}^{*}}\, , \label{eq:neutO}
\end{eqnarray}
in which the gamma-rays emitted upon the deexcitation of oxygen are detectable.

DUNE is a neutrino detector to emerge in mid-decade which employs a $40$ kton liquid Time Projection Chamber (TPC) based on liquid argon. The detector is most sensitive to $\nu_e$s via the charged-current reaction with argon (CCAre):
\begin{eqnarray}
\nu_e  + {^{40}{\rm Ar}} \rightarrow e^{-} + {^{40}{\rm K}^{*}}\, ,
\label{eq:CCAr}
\end{eqnarray}
and is capable of detecting several thousands $\nu_e$s from a CCSN at a distance of $10$ kpc. We note that another charged-current reaction:
\begin{eqnarray}
\bar{\nu}_e  + {^{40}{\rm Ar}} \rightarrow e^{+} + {^{40}{\rm Cl}^{*}}\, ,
\label{eq:CCArb}
\end{eqnarray}
is sensitive to $\bar{\nu}_e$, although the event rate is subdominant due to its smaller $\nu_e$ cross section. In this study, we focus only on the CCAre channel and ignore others. We note, however, that other channels may still be useful. For instance, coherent scattering with Ar via neutral-current reactions has the potential to provide information on all flavors of neutrinos, although there remains many technical issues to be studied \citep[see, e.g.,][]{2015arXiv150907739B,2017APS..DNP.EA119N}.

JUNO is a liquid scintillator detector, which is planned for operation in 2021. It will have a volume of $20$ ktons, and will be most sensitive to $\bar{\nu}_e$s via the IBD-p reaction. Although we use only the IBD-p reaction channel in this study, there are other charged-current reactions. These include:
\begin{eqnarray}
&& \hspace{-15mm} \bar{\nu}_e  + {^{12}{\rm C}} \rightarrow e^{+} + {^{12}{\rm B}}\, ,
\label{eq:CCCarbb} \\
&& \hspace{-15mm}  \nu_e  + {^{12}{\rm C}} \rightarrow e^{-} + {^{12}{\rm N}}\, , \label{eq:CCCarb}
\end{eqnarray}
the neutral-current reaction on carbon
\begin{eqnarray}
\nu_i  + {^{12}{\rm C}}  \rightarrow \nu_i  + {^{12}{\rm C}^{*}}\, , \label{eq:neutC}
\end{eqnarray}
and with protons,
\begin{eqnarray}
\nu_i  + p  \rightarrow \nu_i  + p\, . \label{eq:neutp}
\end{eqnarray}
All these reactions can in principle be useful for studying CCSN neutrinos. As discussed in \citet{2002PhRvD..66c3001B}, those reaction channels may play an important role in retrieving the energy spectra for all neutrino flavors, although the systematic errors may be quite large \citep{2019PhRvD..99l3009L}.

IceCube resides in Antarctica, is a few Mtonnes of pure water ice, and is primarily designed to detect $\gtrsim 100$ GeV neutrinos. As pointed out early by \citet{1988ApJ...329..335P,1996PhRvD..53.7359H}, this type of detector is also capable of detecting $\sim$ MeV neutrinos by taking advantage of the large number of optical modules. Indeed, IceCube has many sensors constructed with a lattice of 5160 digital optical modules containing photomultiplier tubes. They are sensitive to Cherenkov photons produced mainly by IBD reactions, indicating that the collective photomultiplier rate will rise once a burst of $\bar{\nu}_e$ courses through the ice volume \citep{2011A&A...535A.109A,2011JPhCS.309a2029K,2019arXiv190807249C}. By virtue of the large volume of the detector, $\sim 10^6$ events are expected to be detected for a CCSN at a distance of $10$ kpc. Although there are some systematic uncertainties in detector sensitivity \citep[see, e.g.,][for more details]{2011A&A...535A.109A}, we assume $100 \%$ efficiency for event detection with a 3.5-Mtonne fiducial volume \citep{2018MNRAS.480.4710S}. In Sec.~\ref{subsec:timevariability}, we discuss the detectability of the spiral SASI in IceCube data with the assumption that the background noise is $1.48 \times 10^{3} {\rm ms}^{-1}$ \citep[see, e.g., ][for more details]{2011A&A...535A.109A,2013PhRvL.111l1104T}.

\begin{figure*}
  \rotatebox{0}{
    \begin{minipage}{1.0\hsize}
        \includegraphics[width=\columnwidth]{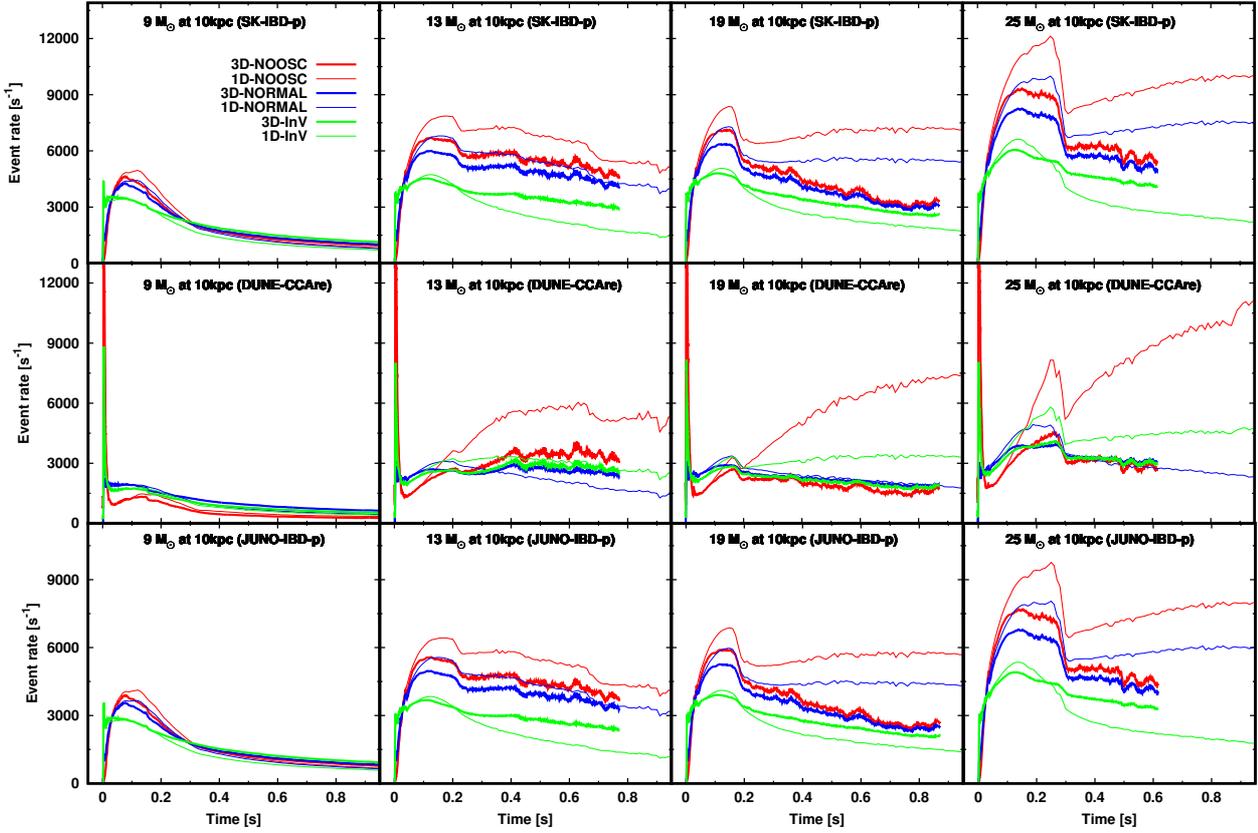}
    \caption{The time evolution of angle-averaged neutrino event rate detected through the major channel for each detector (from top to bottom, SK, DUNE and JUNO) for selected models (from left to right, 9-, 13-, 19- and 25-$M_{\sun}$). The color distinguishes the neutrino oscillation model: no flavor conversions (red), normal mass hierarchy (blue) and inverted mass hierarchy (green). For comparison, we also display 1D counterparts as thin lines. The distance to the CCSN source is assumed to be $10$ kpc.}
    \label{graph_tevoeveratecomparebet1D3D}
  \end{minipage}}
\end{figure*}

\begin{figure*}
  \rotatebox{0}{
    \begin{minipage}{1.0\hsize}
        \includegraphics[width=\columnwidth]{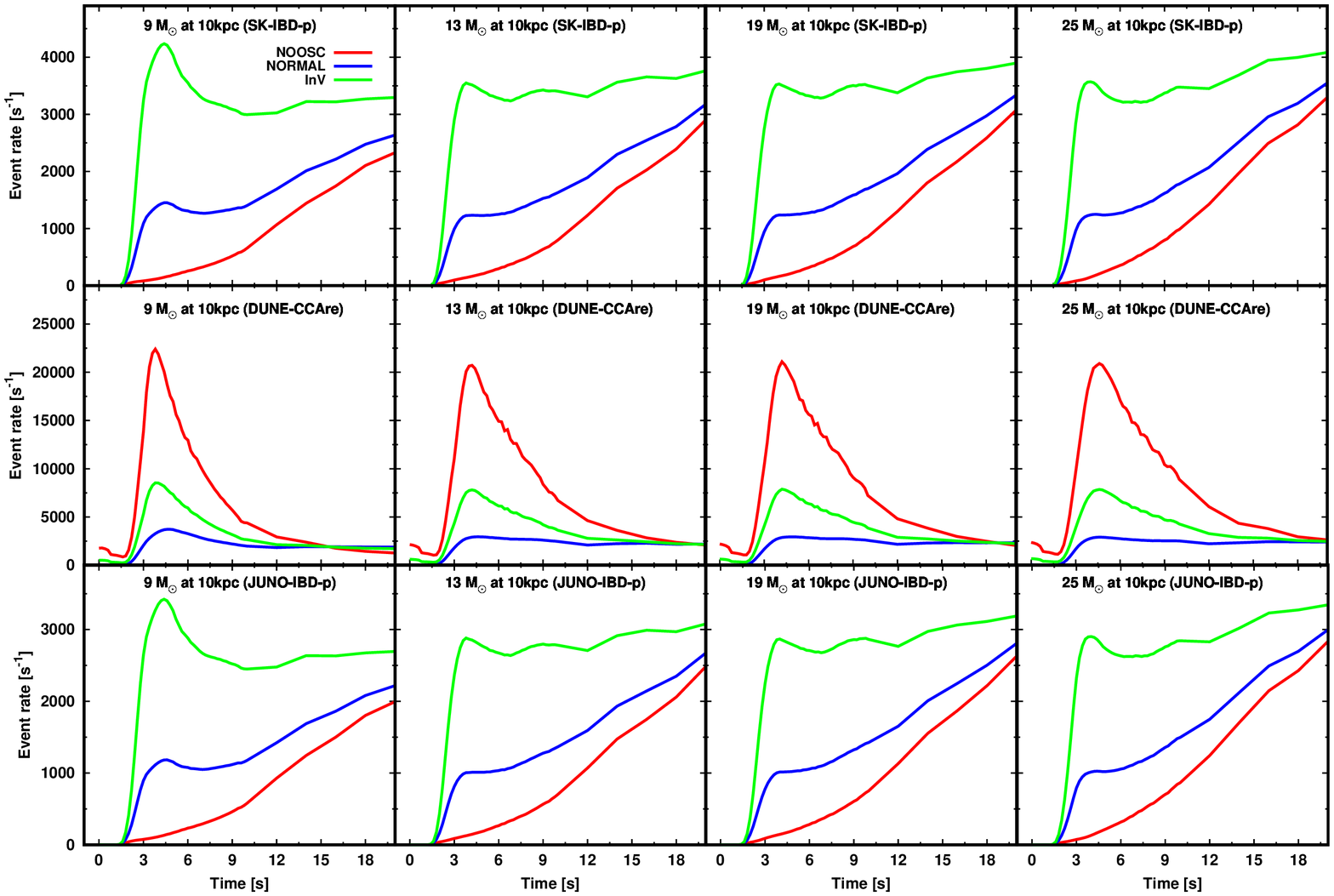}
    \caption{The same as Fig.~\ref{graph_tevoeveratecomparebet1D3D} but focusing only in the earlier phase (${\rm T} < 20$ ms). Since the difference between 1D and 3D is subtle, we only display the result of 1D in this figure.}
    \label{graph_tevoNburst1D_all}
  \end{minipage}}
\end{figure*}

\subsection{Neutrino Oscillations}\label{subsec:neutrinooscillations} 

There has been accumulating experimental evidence for neutrino oscillations, implying that CCSN neutrinos undergo flavor conversion. In this paper, we employ a simplified approach, assuming a purely adiabatic Mikheyev-Smirnov-Wolfenstein (MSW) oscillation model for both normal and inverted neutrino-mass hierarchies. We ignore other possible matter effects, such as non-adiabaticity \citep{1989PhRvD..39.1930K,2000PhRvD..62c3007D} and the effects of Earth's matter \citep{2001NuPhB.616..307L}. Also, we do not distinguish $\nu_{\mu}$, $\nu_{\tau}$, and their anti-particles at the CCSN source in our computations of flavor mixing, an approach which is consistent with our CCSN simulations. However, we distinguish neutrinos and anti-neutrinos at the Earth (see Eqs.~\ref{eq:flavconv_nue}--\ref{eq:flavconv_nuxb} for more details). It should also be mentioned that we do not take into account effects of neutrino-neutrino self-interaction in either our CCSN models or neutrino propagation, although this assumption may not be adequate. The occurrence of fast pairwise flavor conversion in CCSNe has recently been suggested \citep{2019PhRvD.100d3004A,2019ApJ...886..139N,2020PhRvR...2a2046M,2020PhRvD.101b3018D} and extensive studies have been undertaken \citep[see, e.g.][]{2019ApJ...883...80S,2020PhRvD.101d3009J,2020PhRvD.101f3001G,2020JCAP...05..027A,2020arXiv200500459B}. We postpone the incorporation of these effects to the future.

By introducing the neutrino survival probabilities $p$ and $\bar{p}$ for neutrinos and their anti-particles, respectively, the neutrino and anti-neutrino fluxes at the Earth ($F_i$ and $\bar{F}_i$, respectively) can be written as
\citep[see also][]{2000PhRvD..62c3007D}:
\begin{eqnarray}
&&F_{e}  = p  F^{0}_{e}  + \left(1- p  \right) F^{0}_{x}, \label{eq:flavconv_nue} \\
&&\bar{F}_{e}  = \bar{p}  \bar{F}^{0}_{e}  + \left(1- \bar{p}  \right) \bar{F}^{0}_{x}, \label{eq:flavconv_nueb} \\
&&F_{x}  = \frac{1}{2}  \left(1- p  \right) F^{0}_{e} 
+ \frac{1}{2} \left(1+ p  \right) F^{0}_{x}, \label{eq:flavconv_nux} \\
&&\bar{F}_{x}  = \frac{1}{2}  \left(1- \bar{p}  \right) \bar{F}^{0}_{e} 
+ \frac{1}{2}  \left(1+ \bar{p}  \right) \bar{F}^{0}_{x}\, , \label{eq:flavconv_nuxb}
\end{eqnarray}
where $F^{0}_{i}$ and $\bar{F}^{0}_{i}$ denote the neutrino and anti-neutrino fluxes respectively in the case without flavor conversion. The subscript $i$ represents the neutrino flavor, while $x$ denotes either the $\mu$- or $\tau$- neutrino, i.e., $F_{x} = F_{\mu} = F_{\tau}$. Note that we assume in our CCSN models that these heavy leptonic neutrinos and their anti-partners are identical at the source, i.e., $F^{0}_{x} = \bar{F}^{0}_{x}$  (see Sec.~\ref{subsec:3DCCSNmodel}). On the other hand, we distinguish them at the Earth, i.e., $F_{x} \neq \bar{F}_{x}$, since $F^{0}_{e}$ is different from $\bar{F}^{0}_{e}$, implying that heavy leptonic neutrinos and their anti-partners are, in general, no longer identical to each other. As we shall discuss in Sec.~\ref{subsec:retrieving}, it is necessary to distinguish them to retrieve the neutrino spectra at the CCSN source from observed quantities, since the cross-section of the eES reaction is different between neutrinos and their anti-partners. This treatment is different from other previous studies in the literature, including ours \citep{2018MNRAS.480.4710S}.

In the case of the normal mass hierarchy, the survival probabilities can be written as:
\begin{eqnarray}
&& \hspace{-10mm} p = \sin^2 \theta_{13}, \label{eq:p_normal} \\
&& \hspace{-10mm} \bar{p} = \cos^2 \theta_{12} \cos^2 \theta_{13}. \label{eq:barp_normal}
\end{eqnarray}
For the inverted mass hierarchy, they are
\begin{eqnarray}
&& \hspace{-10mm} p = \sin^2 \theta_{12} \cos^2 \theta_{13}, \label{eq:p_inv} \\
&& \hspace{-10mm} \bar{p} = \sin^2 \theta_{13}\, . \label{eq:barp_inv}
\end{eqnarray}
In this study, we adopt the neutrino mixing parameters, $\theta_{12}$ and $\theta_{13}$ from \citet{2017PhRvD..95i6014C}: $\sin^2 \theta_{12} = 2.97 \times 10^{-1}$ and $\sin^2 \theta_{13} = 2.15 \times 10^{-2}$.

\section{Results}\label{sec:result}

\subsection{Progenitor dependence}\label{subsec:ProgenitorDependence}

Unless otherwise stated, the central results we present in this paper assume a CCSN source distance of $d=10$ kpc.
Figure~\ref{graph_tevoeveratecomparebet1D3D} displays the time evolution of the angle-averaged event rate in the major channel of each detector: IBD-p for SK and JUNO, and CCAre for DUNE. We select four representative 3D models (9, 13, 19 and 25 $M_{\sun}$) to highlight some qualitative trends with progenitor. For comparison, we also show the results of their corresponding 1D counterparts as thin lines, and the color distinguishes neutrino oscillation models.

We confirm that DUNE has a unique sensitivity to the neutrino-mass hierarchy through the detection of neutronization burst (${\rm T} \lesssim 20$ms), which is less dependent on progenitor (as suggested previously, see also Fig.~\ref{graph_tevoNburst1D_all}). Up to the ${\rm T} \sim 100$ms, there are no remarkable differences between 1D and 3D, and the progenitor dependence is modest. However, by this time hydrodynamic instabilities have already emerged in the post-shock region. Prompt convection occurs at ${\rm T} \lesssim 40$ms, which provides major perturbations for neutrino-driven convection \citep[see, e.g.,][]{2018ApJ...854..136N}, and non-radial motions behind the shock wave rapidly grow (see \citet{2019MNRAS.490.4622N,2020MNRAS.491.2715B} for more details). Nevertheless, our results suggest that multi-dimensional instabilities in the early postbounce phase have less influence on the neutrino signals. This is mainly attributed to two facts: 1) the angle-averaged hydrodynamic quantities in 3D CCSN models are not so much different from those in 1D and 2) most of the convective region resides in optically thick regions, suggesting that multi-dimensional effects on neutrino signals are masked. We note, however, that low-energy neutrinos ($\lesssim 10$ MeV) can escape from the convective region due to their smaller interaction cross sections. This indicates that multi-dimensional effects can be investigated using the few-MeV neutrinos, although it is not easy to detect them (since those neutrinos are also transparent to the detector medium). The improved sensitivity in liquid scintillator detectors such as Borexino \citep{2009NIMPA.600..568B}, KamLAND \citep{2016ApJ...818...91A}, SNO+ \citep{2015arXiv150805759S}, and JUNO for the few-MeV neutrinos may help with such an analysis. SK may also be capable at these lower energies \citep{2019ApJ...885..133S}, although it would suffer from the low statistical significance.

At $T \gtrsim 100$ ms, on the other hand, remarkable progenitor- and dimensional dependences emerge. Regardless of detector or neutrino oscillation model, we find that the event rate is lowest and highest for the 9 and 25 $M_{\sun}$ progenitors, respectively. This mainly reflects their different accretion histories. Indeed, the neutrino luminosities and event rates for different CCSN models are clearly correlated with mass accretion rate (see Fig.~\ref{graph_Massaccretion_neutsig}). This trend is consistent with previous studies based on 1D CCSN models \citep[see e.g.][]{2019ApJ...881..139S}. We note, however, that the detailed progenitor dependence can not be captured adequately by 1D models, and we discuss this below.

\begin{figure*}
  \rotatebox{0}{
    \begin{minipage}{1.0\hsize}
        \includegraphics[width=\columnwidth]{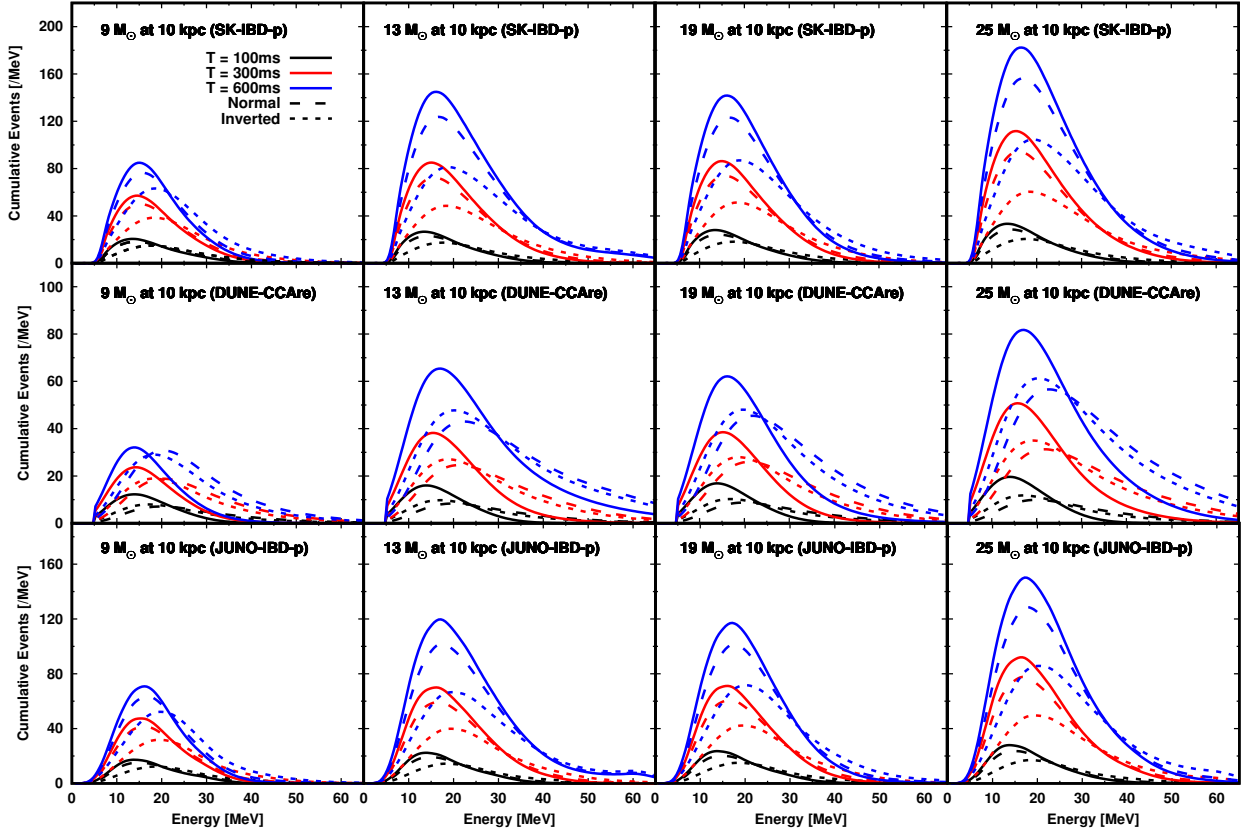}
    \caption{Energy spectrum of time-integrated (cumulative) event counts at three selected times, ${\rm T} = 100$ (black), $300$ (red), and $600$ ms (blue), in the major detection channel for each detector (from top to bottom, SK, DUNE and JUNO) boasting neutrino energy sensitivity and for selected progenitors (from left to right, 9-, 13-, 19- and 25 $M_{\sun}$). Color denotes the time and the line type distinguishes the neutrino oscillation model. The event total is angle-averaged and the distance to the CCSN source is assumed to be $10$ kpc.}
    \label{graph_CumSpect_Pdepe_lastsnapshot}
  \end{minipage}}
\end{figure*}

We start with SK and JUNO results. As described in Sec.~\ref{subsec:neutrinooscillations}, these detectors are most sensitive to $\bar{\nu}_e$ neutrinos (at the Earth). We find that the event rate in 1D models is systematically higher than in 3D in the case with no flavor conversion (red lines in Fig.~\ref{graph_tevoeveratecomparebet1D3D}). One might think that this is simply due to the failure of shock revival for 1D models in which the accretion component of neutrino luminosity is higher and the event rate is increased accordingly. This explanation is qualitatively true, but is not enough to explain the difference. For instance, we find that the 13-$M_{\sun}$ model in 3D has a smaller event rate than that in 1D, in spite of the failure of shock revival for both cases. Where does the difference come from?

A clue to understanding the difference is related to the average energy of $\bar{\nu}_e$ neutrinos at the source, which is shown in the right and middle panels of Fig.~\ref{graph_neutrino_lumi_aveene_neutsig}. As shown, the average energy is systematically higher in 1D than 3D, which results in higher event rates for $\bar{\nu}_e$ neutrinos in SK and JUNO. As already mentioned in Sec.~\ref{subsec:3DCCSNmodel}, PNS convection affects neutrino emission through its effect on the neutrinospheres.  Due to PNS convection, the angle-averaged neutrinospheres in 3D are located at larger radii and lower matter temperatures than those in 1D. The latter effect reduces the average energy of $\bar{\nu}_e$ neutrinos in 3D, and this accounts for much of the difference in neutrino detection rates. The same trend can also be seen for event rates in DUNE (sensitive to $\nu_e$ neutrinos at the Earth), which is higher in 1D than 3D for no-oscillation models (see the middle panel of Fig.~\ref{graph_tevoeveratecomparebet1D3D}).

We now turn our attention to the case of the normal mass hierarchy (blue lines in Fig.~\ref{graph_tevoeveratecomparebet1D3D}), for which the survival probability of $\bar{\nu}_e$ neutrinos is $\sim 70 \%$. This indicates that the characteristics of $\bar{\nu}_e$ neutrinos at the CCSN source are still responsible for the major trends in the event rates at SK and JUNO. We find that the difference in the event rate between 1D and 3D models follows the same trend (however modest) as found in the case with no flavor conversions. On the other hand, the survival probability of $\nu_e$ neutrinos for the normal mass hierarchy is almost zero, implying that the event rates at DUNE reflect the characteristics of $\nu_x$ neutrinos at the CCSN source. Since the $\nu_x$ luminosity is lower than that for $\nu_e$ neutrinos at the CCSN source, the event rate becomes smaller than with no flavor conversions. We find that there are no remarkable differences in the event rates between 1D and 3D for all progenitor models at DUNE. At first glance, this looks a bit strange, since the luminosity of $\nu_x$ neutrinos at the source is systematically higher in 3D than in 1D (due to effects of PNS convection), while the average energy is very similar for the two cases (see Fig.~\ref{graph_neutrino_lumi_aveene_neutsig}).

This complexity can be illuminated by studying high-energy $\nu_x$ neutrinos at the source. Some fractions of $\nu_x$s emitted in the vicinity of PNS experience shock acceleration, which creates a non-thermal tail in the energy spectrum (Hotokezaka and Nagakura in preparation). The smaller shock radius that accompanies high mass accretion rates provides conducive conditions for shock acceleration actually realized in our 1D models (except for the 9-$M_{\sun}$ model). Although non-thermal neutrinos have a smaller contribution to luminosity and average energy, they affect the neutrino event rate at terrestrial detectors, perhaps in measurable ways. We emphasize that high-energy neutrino emission in the 9-, 19- and 25-$M_{\sun}$ models in 3D is subtle, since the shock wave is revived and propagates through a low density medium, making neutrino acceleration inefficient. The increase in neutrino event rate due to a non-thermal component in 1D compensates for the reduction in the event rate due to the absence of PNS convection. Hence, the dimensional dependence is not remarkable for these progenitors. On the other hand, the shock wave stalls and is not revived for the 3D model of 13-$M_{\sun}$, i.e., non-thermal neutrinos can contribute to its event rate. As a result, due to PNS convection the event rate is higher in 3D than in 1D\footnote{It should be noted that the contribution of the high-energy component in the 3D model is smaller than that of 1D. This is mainly because the shock radius in the 3D model is slightly larger than in 1D, which reduces the efficiency of shock acceleration.}.

We now remark on the non-thermal component of $\nu_x$ neutrinos. As pointed out by \citet{2008PhRvL.100a1101L}, inelastic neutrino-nucleus scatterings may suppress the high-energy tail of the neutrino spectrum. Our CCSN simulations, however, do not incorporate the inelasticity in this reaction, though the velocity dependence and Doppler effects are included. If the non-thermal component is suppressed, the event rate of 1D models would be systematically smaller than we obtain. We also note that accurate computations of shock acceleration requires high resolution neutrino transport simulations in both real and moment spaces, and neutrino transport with multi-angle treatment is also highly desired. Thus, we stress that there remain many uncertainties in the high-energy tail of $\nu_x$ neutrinos, which will be addressed in the future. It should be noted, however, that those uncertainties do not influence the event rates for our 3D CCSN models, in particular for successful explosions, since the shock alteration of the high-energy tail of the $\nu_x$ spectrum is negligible.

In the case of the inverted mass hierarchy (green lines in Fig.~\ref{graph_tevoeveratecomparebet1D3D}), we find that the difference between 1D and 3D models in the event rates in SK and JUNO are modest. In the oscillation model, the event rate reflects the property of $\bar{\nu}_x$ neutrinos at the CCSN source. As in the case in DUNE for the normal mass hierarchy, the non-thermal component of $\bar{\nu}_x$ neutrinos contributes interestingly to the 1D event rates. It should be noted, however, that the IBD-p cross section is less sensitive to the neutrino energy than is that for the CCAre, implying that the contribution from high-energy components is more modest in SK and JUNO than in DUNE. As a result, the increase of the $\bar{\nu}_x$ luminosities due to PNS convection directly accounts for the higher event rate in SK and JUNO when comparing 3D and 1D, regardless of progenitor. On the other hand, the difference between the 1D and 3D event rates in DUNE can be explained by the same mechanism as in the case with the normal-mass hierarchy. For the inverted mass hierarchy, the survival probability of $\nu_e$ neutrinos is $\sim 30 \%$, implying that $\nu_x$ neutrinos at the source significantly affect the event rate. We find that the event rate is systematically higher in 1D than 3D for the 19 and 25 $M_{\sun}$ models, which is mainly due to the contribution of the high-energy $\nu_x$ neutrinos. We emphasize that the CCAre reaction has a higher sensitivity to high-energy neutrinos than does the IBD, with the result that the effects of the non-thermal tail in DUNE may be substantial. We also note that the difference for 13 $M_{\sun}$ model between 3D and 1D is modest, since non-thermal neutrinos are created in the 3D model and the shock fails.

\begin{figure*}
  \rotatebox{0}{
    \begin{minipage}{1.0\hsize}
        \includegraphics[width=\columnwidth]{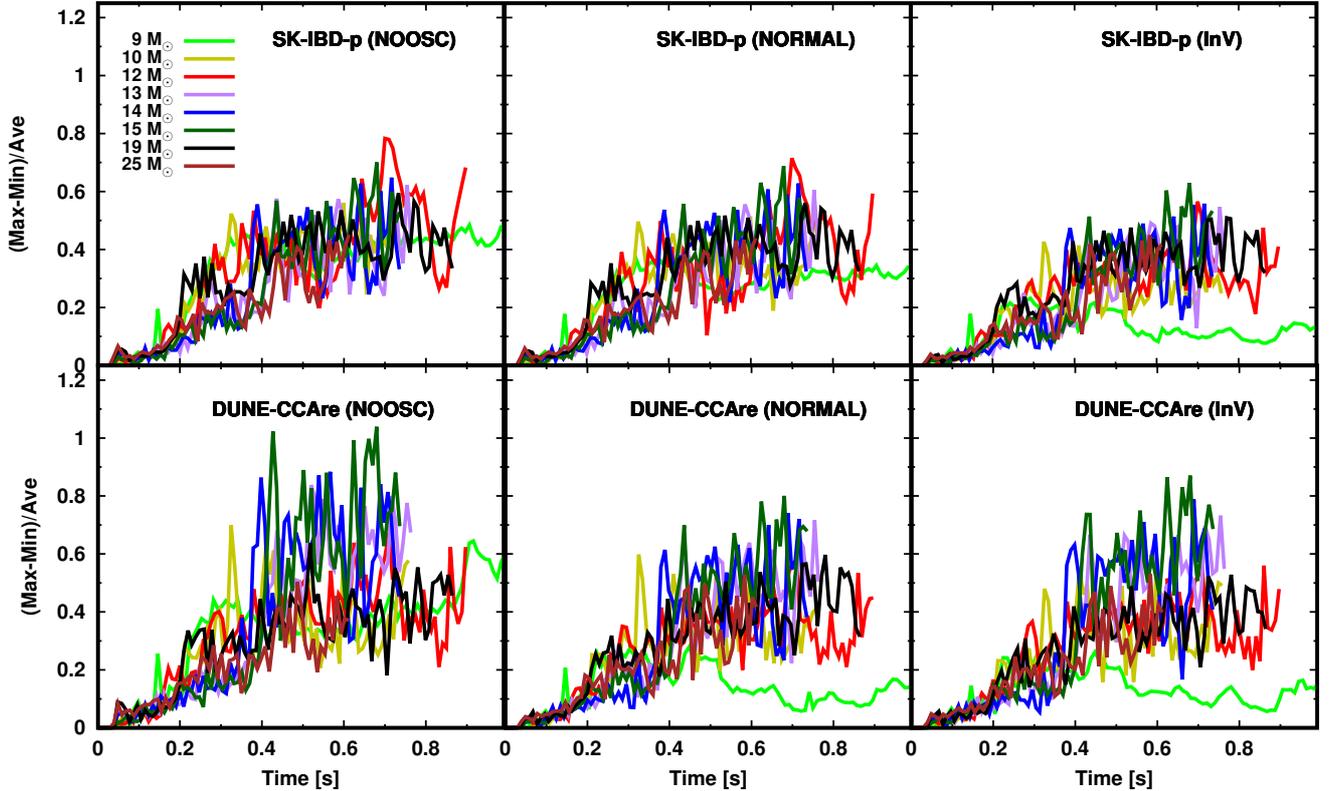}
    \caption{Angular variation of the neutrino event rate at SK (top) and DUNE (bottom) as a function of time. The asymmetry is measured by the difference between maximum and minimum event rate normalized by the angle-averaged one. Color distinguishes CCSN models with different progenitor masses.}
    \label{graph_Tevo_AsymOn}
  \end{minipage}}
\end{figure*}

\begin{figure*}
  \rotatebox{0}{
    \begin{minipage}{1.0\hsize}
        \includegraphics[width=\columnwidth]{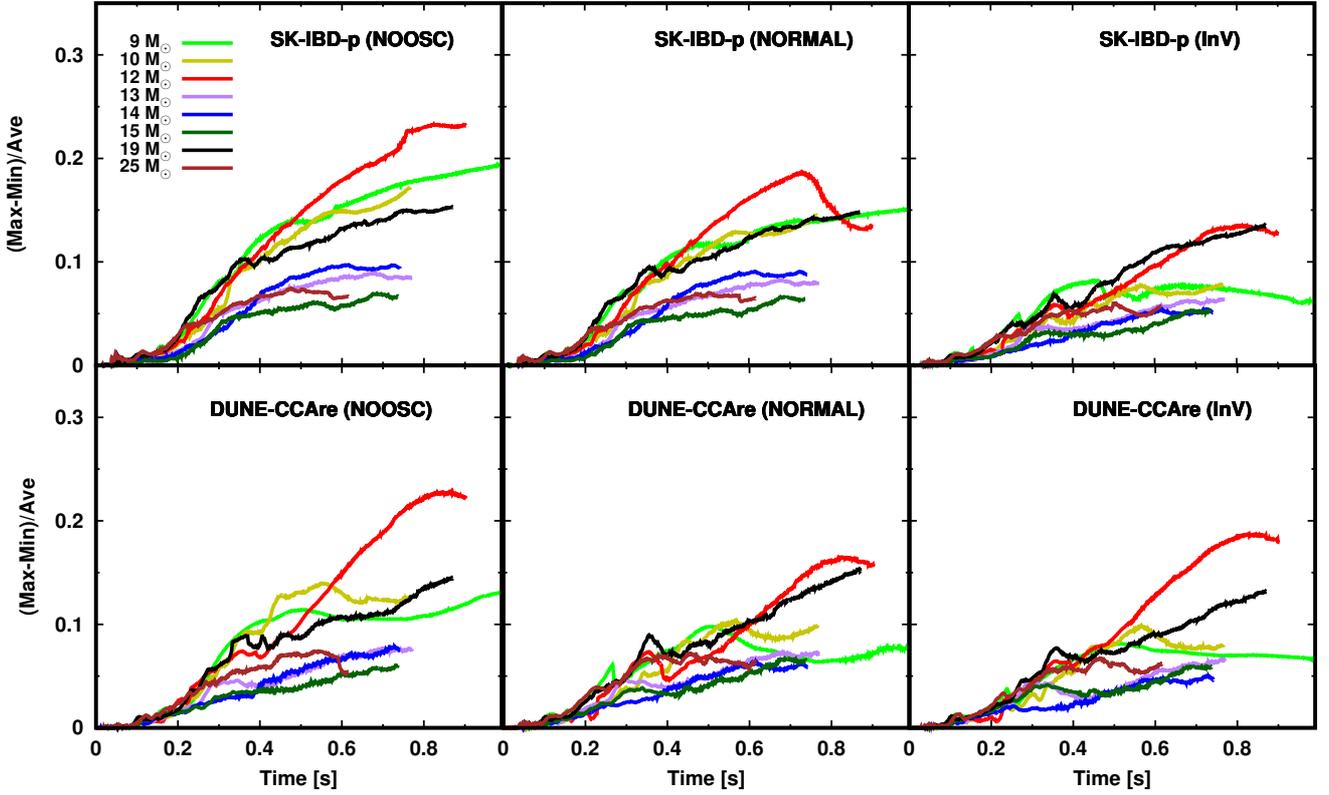}
    \caption{Same as Fig.~\ref{graph_Tevo_AsymOn}, but for the time-integrated (cumulative) event rate.}
    \label{graph_Tevo_AsymCum}
  \end{minipage}}
\end{figure*}

\begin{figure*}
  \rotatebox{0}{
    %\begin{minipage}{1.0\hsize}
    \begin{minipage}{0.8\hsize}
        \includegraphics[width=\columnwidth]{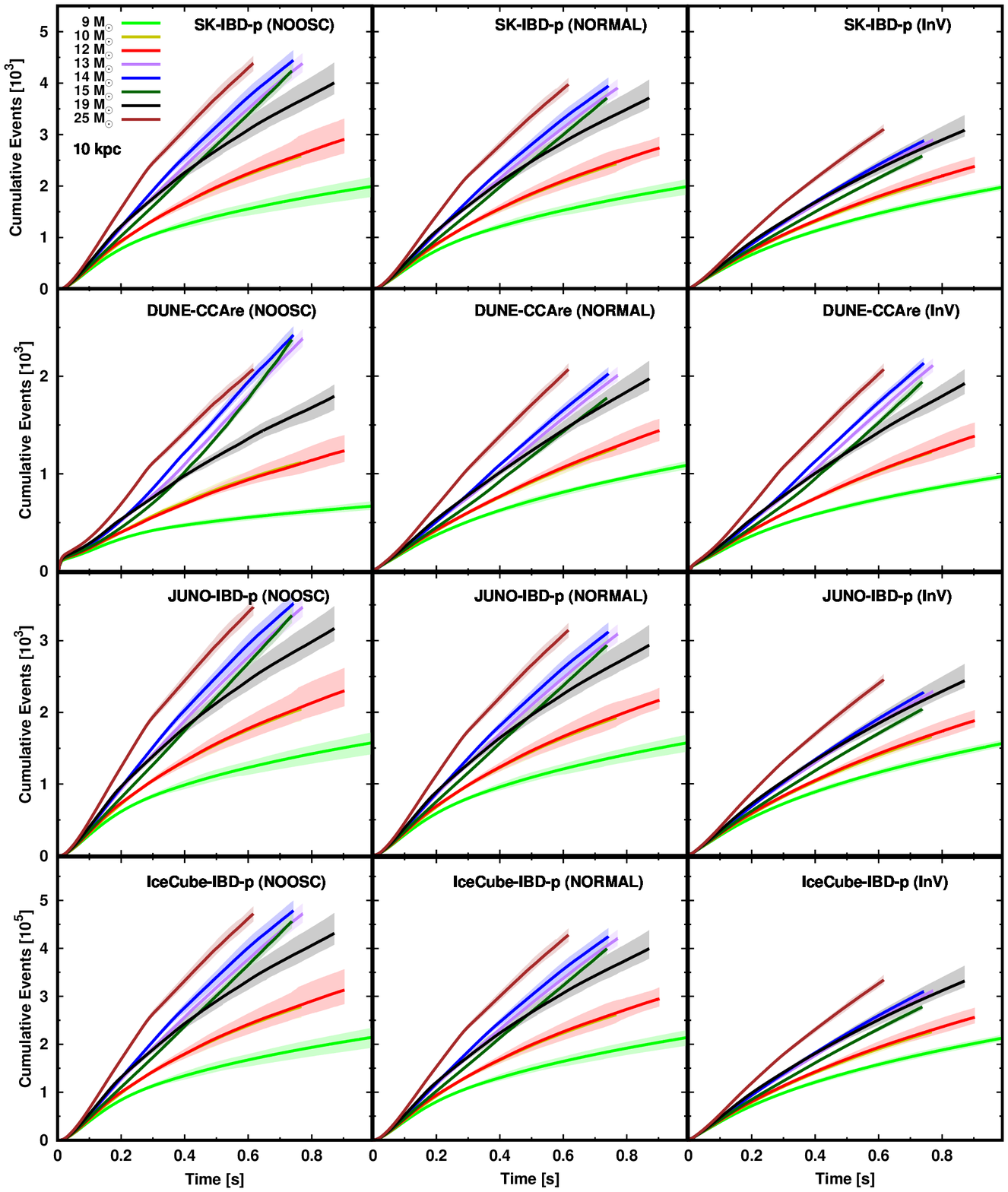}
    \caption{Time evolution of the cumulative number of events in the major channel of each detector (from top to bottom, SK, DUNE, JUNO, and IceCube). Left: no flavor conversions. Middle: normal mass hierarchy. Right: inverted mass hierarchy. Color distinguishes the 3D models. The angular variations are also displayed with the shaded region along each solid line and the solid line corresponds to the angular average. The distance to the CCSN is assumed to be $10$ kpc.}
    \label{graph_Tevo_Cumu}
  \end{minipage}}
\end{figure*}

Summarizing the above results, the details of the progenitor dependence and the difference between 1D and 3D models are determined by a complex interplay. At SK and JUNO, the event rate in 1D is systematically overestimated (underestimated) for the case with the normal (inverted) mass hierarchy. For DUNE, the event rate is overestimated for the 1D case with the inverted mass hierarchy, while the dimensional differences are not remarkable for the normal mass hierarchy. It is important to point out that 1D models overestimate the progenitor- and neutrino oscillation dependence of the event rate for all detectors. The former trend is mainly due to the diminished accretion-power in 3D models at later post-bounce times. The latter is due mainly to effects of PNS convection, for which the increase in the $\nu_x$ luminosity and the decrease in the average energy of $\nu_e$ and $\bar{\nu}_e$ neutrinos reduces the flavor dependence of the event rate in all detectors. We find that the variation due to different oscillation models for a common 3D CCSN model is within a factor of $\sim 1.5$ through the end of our simulations.

In Fig.~\ref{graph_CumSpect_Pdepe_lastsnapshot}, we provide energy spectra for the cumulative total number of events for the major channel of each detector for three selected post-bounce times: ${\rm T} = 100, 300$ and $500$ ms. We selected four representative 3D models. In computing these spectra, we took into account the smearing effects described in SNOwGLoBES. Regardless of progenitor, detector, or neutrino oscillation model, the peak of each energy spectrum increases with time and is located between $\sim 15$ and $\sim 20$ MeV. The overall shape of the energy spectra are very similar for different progenitors, except for the 9 $M_{\sun}$ model. This difference is due to the early lack of an accretion component of the neutrino emission, which reduces not only luminosities, but also the average energy of the neutrinos (see  Fig.~\ref{graph_neutrino_lumi_aveene_neutsig}). Furthermore, we find that the spectra become stiff once flavor conversion is taken into account. This is attributed to the fact that both $\nu_e$ and $\bar{\nu}_e$ spectra at the Earth are affected by $\nu_x$ (and $\bar{\nu}_x$) neutrinos at the CCSN source. It should be noted, however, that the total neutrino count in an energy bin decreases with increasing energy ($\gtrsim 30$ MeV), This implies that the signal-to-noise ratio (SN-ratio) is smaller for the high-energy tail and that it would be challenging to measure it.  Thus, it would be better to study a different reaction channel with a higher sensitivity to high-energy neutrinos. Addressing the issue is a beyond the scope of this paper.

\subsection{Asymmetry of neutrino signals}\label{subsec:Asymmetry}

It has been observed in multi-dimensional CCSN simulations that various mechanisms cause asymmetric neutrino emission \citep[see, e.g.][]{2018ApJ...865...81O,2018PhRvD..98l3001W,2019ApJ...872..181H,2019PhRvD.100f3018W,2019ApJ...880L..28N,2019ApJ...881...36G,2019MNRAS.489.2227V}. This implies that the neutrino event rates depend on the angular location of the observer. Therefore, it is necessary to quantify the magnitude of such variations in the event rates at detectors. In this section, we quantify this quantity for our 3D CCSN models.

In our CCSN simulations, we employ $128 \times 256$ ($\theta \times \phi$) angular grid points; hence, we can in principle compute the neutrino event rate in each angular direction. However, those computations are very expensive and we are less interested in the results at a specific angular point, but rather in the global characteristics. For this purpose, we perform a spherical harmonic decomposition of the neutrino data obtained from our CCSN simulations. In each energy bin of the first-moment of the neutrinos, we carry out the spherical harmonics decomposition by following Eqs.1-3 in \citet{2012ApJ...759....5B}. We then compute the neutrino event rate by using SNOwGLoBES for each spherical harmonic coefficient, and then merge the results up to $\ell = 2$, taking into account the azimuthal dependence. Although this approach may discard fine angular structure of the neutrino signal, those structures are very sensitive to the realization of our numerical CCSN models. This indicates that statistical analyses with many realizations are required to study the fine structure, which is beyond the scope of this paper. Nevertheless, our analysis captures the overall trends of the angular distributions of the neutrino event rates.

Figure~\ref{graph_Tevo_AsymOn} portrays the angular asymmetry in the neutrino detection rate at SK and DUNE as a function of time. Since they are almost the same as in the case of SK, we omit the corresponding JUNO and IceCube plots. We define the degree of asymmetry degree as the difference between the maximum and minimum of the neutrino event rate, normalized by the angular average. Roughly speaking, the degree of asymmetry increases with time, except in the case of the 9-$M_{\sun}$ progenitor in both neutrino oscillation models. This exception is attributed to a combination of effects; 1) the earlier lack of an accretion component reduces the contribution of $\bar{\nu}_e$ neutrinos to the signal, implying that the $\nu_x$ neutrino contribution is greater in the 9-$M_{\sun}$ model, and 2) $\nu_x$ neutrino emissions are less asymmetric than other species (this trend is common for other CCSN models as well). As a result of these two effects, the asymmetry of the neutrino event rate is smaller.

For other progenitors, the degree of asymmetry exceeds $\sim 50 \%$ for SK. Meanwhile, it reaches $\sim 100 \%$ for DUNE (see Fig.~\ref{graph_Tevo_AsymOn}). As shown in this figure, the strong asymmetry in DUNE is observed only for non-exploding models (13-, 14- and 15-$M_{\sun}$), indicating that this is associated with the spiral SASI. As pointed out by \citet{2014ApJ...786..118I}, the spiral SASI tends to emerge when the stalled shock wave is located at smaller radii ($\lesssim 100$ km), which is realized in our non-exploding models \citep{2020MNRAS.491.2715B}. Our result also suggests that $\nu_e$ neutrino emissions at the source have the highest asymmetry, which also reflects the impact of the spiral SASI \citep[see also][]{2019MNRAS.489.2227V}. The spiral SASI effect is prominent in the vicinity of the shock wave, which affects the accretion component for $\nu_e$ neutrinos substantially. Although $\bar{\nu}_e$ neutrinos are also influenced by the spiral SASI, their corresponding neutrinospheres are located at smaller radii than those for $\nu_e$ neutrinos, indicating that the impact of the spiral SASI for them becomes modest. We also find that the degree of asymmetry for neutrino oscillation models tends to be smaller than that for models with no flavor conversions, and this is due to the less asymmetric emissions of $\nu_x$ (and $\bar{\nu}_x$) neutrinos at the source.

Although there is a strong angular dependence in the instantaneous neutrino event rates, the angular asymmetry in the time-integrated count numbers tends to be more modest.  Indeed, the asymmetry in the cumulative number of events is less than $\sim 20 \%$ among all our models (see Fig.~\ref{graph_Tevo_AsymCum}). We also see a progenitor dependence in the difference between the event rate and the cumulative number of events. For non-exploding models, the asymmetry in the cumulative number of events is substantially reduced by integrating over time. This is attributed to the fact that the asymmetric neutrino emissions induced by the motions of the spiral SASI partially average out, smoothing the angular profile. On the other hand, the asymmetry of the 9-$M_{\sun}$ model behaves in an opposite way to that of the non-exploding models. As already pointed out, the asymmetry of this event rate is the smallest among our models, though the cumulative number of events is comparable to that for other models. This is attributed to the fact that the asymmetric neutrino emission in the 9-$M_{\sun}$ model seems to be mainly associated with the LESA, in which anti-correlated asymmetry between $\nu_e$ and $\bar{\nu}_e$ neutrino emissions appear coherently. Because of this coherency, the angular distribution of the cumulative number of events is not smoothed by the time integration, but rather accumulates with time. The asymmetric neutrino emissions of the 12-$M_{\sun}$ model is also associated with the LESA \citep[see, e.g.,][]{2019MNRAS.489.2227V}. Hence this model has the highest asymmetry in the cumulative number of events among our models.

Fig.~\ref{graph_Tevo_Cumu} portrays the time evolution of cumulative number of events including its angular dependence. The shaded region corresponds to the range of the angular variations. We find that the angular variation is not large enough to smear out the progenitor dependence, implying that we may be able to place constraints on CCSN progenitors from the time evolution of the cumulative number of events. We also note that the uncertainty in the angular variations is comparable to the anticipated Poisson noise over a time integration with a window of $\sim 20$ms, assuming the source distance is at $10$ kpc. This indicates that Poisson noise does not compromise the constraint on the progenitor star mass using the time evolution of the cumulative number of events for ${\rm T} \gtrsim 100$ms.

\subsection{Time variability due to the Spiral SASI}
\label{subsec:timevariability}

\begin{figure*}
  \rotatebox{0}{
    \begin{minipage}{0.8\hsize}
    %\begin{minipage}{1.0\hsize}
        \includegraphics[width=\columnwidth]{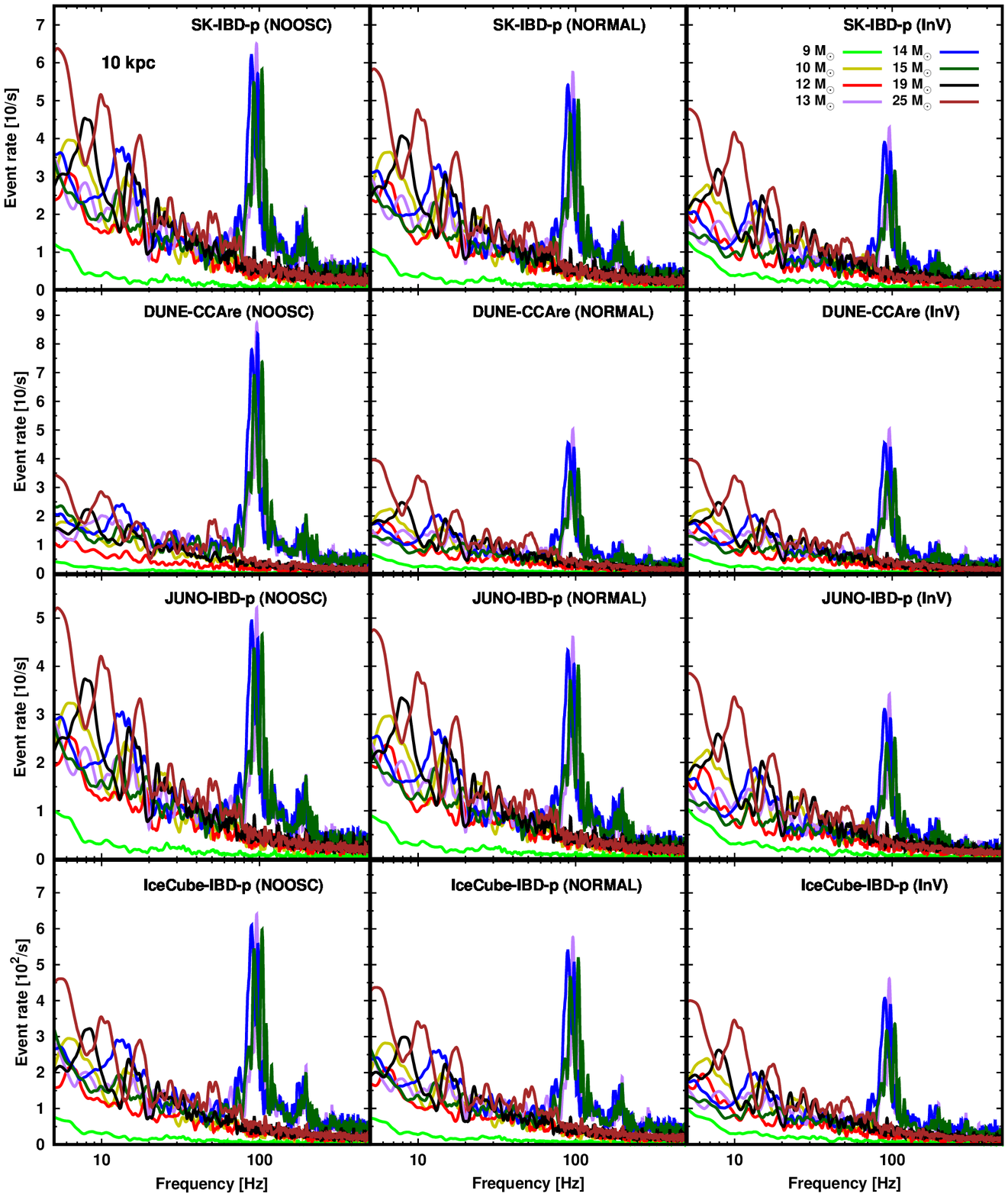}
    \caption{The finite-time Fourier transform of the neutrino event rate in each detector (from top to bottom, SK, DUNE, JUNO, and IceCube). Left: no flavor conversions. Middle: normal mass hierarchy. Right: inverted mass hierarchy. Color distinguishes the 3D models. The distance to the CCSN source is assumed to be $10$ kpc. A remarkable sharp peak at $\sim 100$ Hz, which corresponds to the temporal behavior induced by the spiral SASI, is seen, but only for non-exploding models. See the text for details.}
    \label{graph_TimeFourier}
  \end{minipage}}
\end{figure*}

As discussed above, the spiral SASI commonly occurs in non-exploding models (13-, 14- and 15-$M_{\sun}$) and this is a qualitatively distinct characteristic from the exploding models. In this section, we discuss the temporal behavior of the neutrino signals in more detail.

In Fig.~\ref{graph_TimeFourier}, we show the Fourier transform of the neutrino event rate in each detector. In the computation,  we apply a Hann window function to the event rate during the time from core bounce to the end of each simulation. We also subtract the angle-averaged event rate to clearly see the spiral SASI feature\footnote{Note that the angle-averaged neutrino luminosity in our 3D models artificially fluctuates due to grid noises, which smears the SASI feature in the Fourier spectrum.}. Note also that we chose twenty-four sight lines and then take the average over the spectra. As shown in these plots, the temporal behavior of the spiral SASI is clearly encoded in the neutrino signals; a strong peak can be seen at $f \sim 100$ Hz ($f$ denotes the frequency) and a modest peak is also observed at $f \sim 200$ Hz in all detectors. We also find that the peak is weaker in neutrino oscillation models than in the case with no flavor conversions, and we attribute this to the fact that $\nu_x$ neutrinos are less sensitive to the spiral SASI than $\nu_e$ and $\bar{\nu}_e$ neutrinos. This is consistent with the findings in our previous paper \citep{2019MNRAS.489.2227V}.

However, it is a challenging to resolve the temporal behavior of the spiral SASI in neutrino signals. Below, we describe the degree of difficulty and discuss what is necessary to resolve it in each detector. We define $\alpha$ as a ratio of the neutrino energy associated with the spiral SASI (corresponding to the region around $f \sim 100$ Hz in our models) to the total emitted energy. According to previous studies \citep[see also][]{2012PhRvD..86j5031L,2013PhRvL.111l1104T,2018PhRvD..98l3001W,2019MNRAS.489.2227V}, $\alpha$ is roughly a few percent or less. Thus, we set $\alpha = 10^{-2}$ as a representative value. We also define another variable, $\Delta {\rm T}_{\rm SASI}$, which denotes the duration of the spiral SASI. Although it depends on progenitor model, we set $\Delta {\rm T}_{\rm SASI}=500$ ms as a typical value. By using these values, we can estimate a number of neutrino events ($N_{\rm signal}$) associated with the spiral SASI for a duration of $1/f$:
\begin{eqnarray}
N_{\rm signal} = 2  
\left( \frac{N_{\rm 10 kpc}}{10^{4}} \right)
\left( \frac{\alpha}{0.01} \right)
\left( \frac{f}{100~{\rm Hz}} \right)^{-1} \nonumber \\
\left( \frac{\Delta {\rm T}_{\rm SASI}}{500~{\rm ms}} \right)^{-1}
\left( \frac{d}{10~{\rm kpc}} \right)^{-2}\, ,
\label{eq:signalSASI}
\end{eqnarray}
where $N_{\rm 10 kpc}$ denotes the expected number of events if the source is located at 10 kpc; $N_{\rm 10 kpc}=10^4$ corresponds to the case with SK. We can estimate the Poisson noise of the event over the corresponding duration, $1/f$:
\begin{eqnarray}
N_{\rm noise} = 5
\left( \frac{N_{\rm 10 kpc}}{10^{4}} \right)^{0.5}
\left( \frac{f}{100~{\rm Hz}} \right)^{-0.5} \nonumber \\
\left( \frac{\Delta {\rm T}}{4~{\rm s}} \right)^{-0.5}
\left( \frac{d}{10~{\rm kpc}} \right)^{-1}\, ,
\label{eq:noiseSASI}
\end{eqnarray}
where $\Delta {\rm T}$ corresponds to the timescale of the overall neutrino emissions. We set $\Delta {\rm T}=4$ seconds as an example. As can be seen in these equations, the Poisson noise dominates the signal for SK if the CCSN is located at 10 kpc\footnote{We can estimate the signal and noise for DUNE and JUNO using Eq.~\ref{eq:noiseSASI} by changing $N_{\rm 10 kpc}$. Note that we arrive at a similar conclusion.}. If we require that the signal-to-noise ratio (SN-ratio) is larger than 5, then the source should be located very nearby ($d \lesssim 1$ kpc). For HK, the threshold distance for HK is still $\sim 2$ kpc\footnote{We assume a fiducial volume of 220 ktons in HK for CCSN neutrino analysis \citep{2018arXiv180504163H}.}.

However, \citet{2013PhRvL.111l1104T} claim that HK is capable of detecting the temporal variation of the SASI in CCSNe at $d \ge 10$ kpc. There are some reasons for this inconsistency. First, those authors assume that the available volume in HK is 740 ktons, which is more than three times larger than we assume. Second, their conclusion hinges on the choice of the most optimal observer direction in which the SASI modulation is strongest. Our conclusions rest on the use of the angle-averaged modulations, so their choice enhances the detectability of the temporal modulation in HK. We also note that in their analysis the SN-ratio required to catch the temporal behavior seems to be smaller than ours. They might be setting $\sim 2$ \citep[see the bottom panel of Figure 1 in][]{2013PhRvL.111l1104T}. Since the SN-ratio is proportional to $d^{-1}$, the threshold distance becomes 2.5 times larger than ours by the difference in the SN-ratio required for detection. As such, their conclusions are based on optimistic choices, and this accounts almost completely for the different conclusions.

For IceCube, on the other hand, the expected number of events is two orders of magnitude larger than in SK, suggesting that it may be the most promising neutrino detector for investigating the temporal behavior due to the spiral SASI \citep[see also][]{2013PhRvL.111l1104T,2018PhRvD..98l3001W}. However, as mentioned, the noise characteristics of IceCube are different from those in other detectors, and this should be taken into account in the discussion of  detectability. Here, we consider this following \citet{2013PhRvL.111l1104T}.

From Eq.~\ref{eq:signalSASI}, we obtain $N_{\rm signal}=200$ for a CCSN at a distance of $d=10$ kpc\footnote{In our simulations, we find that $N_{\rm signal}$ is $\sim 500$ (see the bottom panels of Fig.~\ref{graph_TimeFourier}), which is consistent to a factor of two with our rough estimate.}. On the other hand, the background noise can be estimated as $1.48 \times 10^6 \times (1/100) = 1.48 \times 10^4$, implying that it is significantly larger than $N_{\rm signal}$. However, the background noise is nearly in steady state. Hence, its time-averaged value can be subtracted. After the subtraction of the time-averaged component, the residual noise is the Poisson noise of the background and CCSN signals, estimated as:
\begin{eqnarray}
N_{\rm noise (IC)} = 
\left( \frac{1.48 \times 10^{6}}{f} + \frac{N_{\rm 10 kpc}}{\Delta {\rm T} f} (\frac{d}{10~{\rm kpc}})^{-2}  \right)^{0.5}\, ,
\label{eq:noiseSASI_IceCube}
\end{eqnarray}
which yields $N_{\rm noise (IC)} \sim 130$ for $\Delta {\rm T}=4$~s and $d=10$~kpc, indicating a SN-ratio of $\sim 2$. We note that our 3D models have a $\sim 2.5$ times higher signal than the rough estimation, indicating that the SN-ratio is $\sim 5$\footnote{Note that the background noise is dominant at $d=10$ kpc. Hence, the SN-ratio is roughly proportional to the signal.}. Thus, our results suggest that the temporal behavior of the spiral SASI at 10 kpc may be seen in the neutrino IceCube signal, albeit with low statistics. We note that, if the CCSN source distance is $d \lesssim 3$~kpc, the temporal fluctuation of the noise is dominated by the Poisson noise of the CCSN neutrinos (not the background), indicating that Eq.~\ref{eq:noiseSASI} becomes a good approximation to Eq.~\ref{eq:noiseSASI_IceCube}. The SN-ratio becomes $\gtrsim 10$ at a distance of $d \lesssim 3$~kpc. Thus, we conclude that IceCube is currently the best detector with which to study the temporal behavior of the spiral SASI, a conclusion consistent with the results of previous studies \citep{2013PhRvL.111l1104T,2020PhRvD.101l3028L}.

\subsection{Correlations}\label{subsec:correlations}

\begin{figure*}
  \rotatebox{0}{
    %\begin{minipage}{1.0\hsize}
    \begin{minipage}{0.8\hsize}
        \includegraphics[width=\columnwidth]{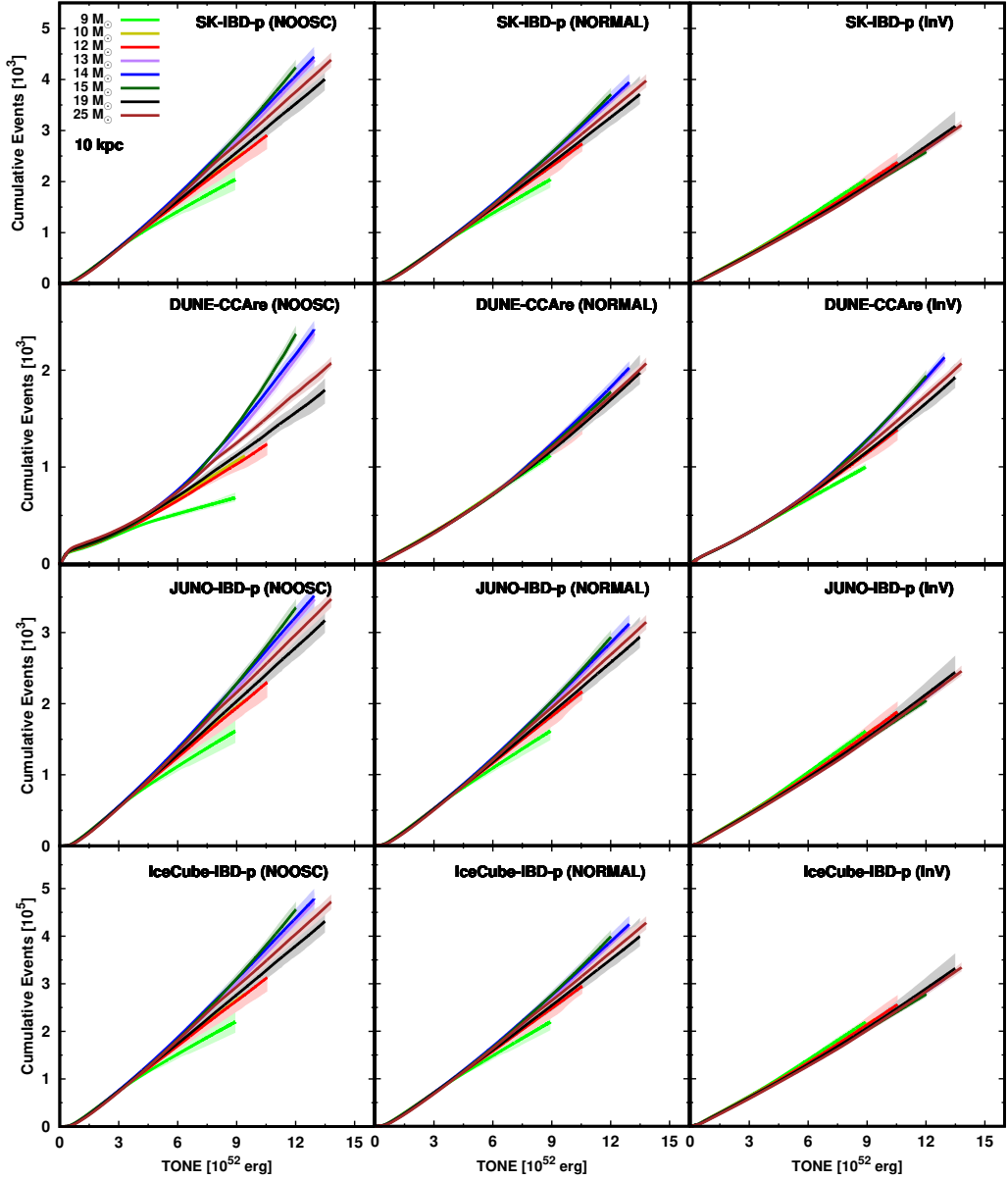}
    \caption{Same as Fig.~\ref{graph_Tevo_Cumu}, but as a function of the total neutrino energy (TONE).}
    \label{graph_EtotvsCum}
  \end{minipage}}
\end{figure*}

Once we actually capture neutrino signals from CCSNe, we will attempt to estimate the total emitted energy of the neutrinos. This is directly related to the neutron star binding energy, a function of the residue mass and nuclear EOS. The most straightforward approach is to reconstruct the energy spectra for all flavors of neutrino at the Earth and then to compute the total radiated energy, taking into account the distance to the CCSN. As we shall see in Sec.~\ref{subsec:retrieving}, it is possible to use the data from multiple detectors to optimize the scientific return. However, the accuracy of the spectrum reconstruction depends on the SN-ratio of the neutrino signal and a reliable estimation is possible only if the source is nearby. A major limitation in this process is also due to the estimates for heavy lepton neutrinos, detected only through reaction channels with moderate sensitivity. As a result, Poisson noise smears the energy spectra and induces large errors in estimates of the total radiated neutrino energy.

In this section, we consider this problem from a different angle, in which we incorporate the results of theoretical models in the analysis. More specifically, we look for theoretical correlations between observed quantities (e.g., the number of events at each detector) and the total neutrino energy (TONE) based on our 3D models. Although this approach is affected by uncertainties in the input physics in our CCSN models, it can be compared with the estimations based on direct spectrum reconstructions from observed data. We, hence, propose this method to combine with theoretical models as a complementary method to those based on purely observed quantities. It should also be emphasized that this approach can be very powerful, in particular for the analyses of CCSNe at great distances, since we would need only the energy- and time-integrated total event numbers in the major detection channel for each detector.

In advance of delving into the analysis, we mention an important point. In the previous sections, we revealed that the neutrino signals have a rich diversity across the progenitor continuum. Indeed, the time evolution of both event rate (see Fig.~\ref{graph_tevoeveratecomparebet1D3D}) and cumulative event number (see Fig.~\ref{graph_Tevo_Cumu}) strongly depends on the progenitor. This could be a major obstacle to our proposed method, since we do not know a priori the CCSN progenitor. Hence, it might be necessary to find correlations which are less sensitive to the progenitor\footnote{However, a galactic supernova will be a target of a vast array of telescopes and the progenitor star, its mass, and distance are likely to be strongly constrained using photon observations.}.

Similar to Fig.~\ref{graph_Tevo_Cumu}, we provide the cumulative number of events at each detector in Fig.~\ref{graph_EtotvsCum}, but as a function of TONE. Note that the TONE monotonically increases with time; thus, the map between time and TONE is monotonic for each model. For neutrino oscillation models, we find that TONE is less sensitive to progenitor. We also find that the uncertainty due to the progenitor dependence is comparable to that of the angular dependence (shaded region), i.e., the error is within a few tens of percent. Hence, we conclude that the correlation is nearly universal (with little progenitor dependence). We emphasize that such universality is not trivial, since the reaction channels used in this study are not sensitive to heavy lepton neutrinos at the Earth. Nevertheless, our result suggests that the cumulative number of events in each reaction channel is universally correlated with the TONE.

As shown in Fig.~\ref{graph_EtotvsCum}, the progenitor dependence of the correlation is weaker for neutrino oscillation models vis \`a vis the no-oscillation models. One of the reasons is that the total radiated energy of the four heavy lepton neutrinos ($\nu_{\mu}$, $\nu_{\tau}$ and their anti-partners) constitutes the dominant contribution to the TONE, although the individual contributions are smaller than for the $\nu_e$ and $\bar{\nu}_e$ neutrinos\footnote{This is mainly due to the absence of charged-current reactions for heavy lepton neutrinos in supernova matter.}. This fact indicates that $\nu_x$ neutrinos at the source contain the most important information concerning the radiated total energy. For no-oscillation models, however, the observed data in the major reaction channels for each detector do not reflect $\nu_x$ neutrino properties at all, and as a result they tend to be less sensitive to the total energy. It should be mentioned that the ratio of the emitted energy of $\nu_e$ or $\bar{\nu}_e$ neutrinos to $\nu_x$ neutrinos varies with progenitor\footnote{This is due to the fact that the accretion component of the neutrino luminosity, which is the dominant contribution in $\nu_e$ and $\bar{\nu}_e$ emission, varies with different progenitor models.} (see also Fig.~\ref{graph_neutrino_lumi_aveene_neutsig}). This results in a large progenitor dependence in the correlation. For oscillation models, on the other hand, the observed data reflect some characteristics of $\nu_x$ neutrino emission at the source, and the degree of correlation depends upon the survival probability of the neutrinos. For the normal-mass hierarchy, $\nu_x$ neutrinos at the source turn into $\nu_e$ neutrinos at the Earth almost completely, indicating that DUNE would provide the most sensitive data with which to measure the TONE. For the inverted-mass hierarchy, $\bar{\nu}_x$ neutrinos at the source change into $\bar{\nu}_e$ neutrinos at the Earth, indicating that the observed data via the IBD-p reaction channel in SK, JUNO, and IceCube have the most direct correlation with the TONE, all of which is consistent with the results displayed in Fig.~\ref{graph_EtotvsCum}.

For convenience, we provide approximate formulae for the correlation in the case of the neutrino oscillation models. We fit the relation to quadratic functions. The fitting formulae are given in the case of normal mass hierarchy as:
\begin{eqnarray}
&{\rm [SK-IBDp-NORMAL]}& \nonumber \\
&N_{\rm Cum} = \left( 220 \hspace{0.5mm} E_{52} + 5 \hspace{0.5mm} E_{52}^2 \right)
\left(  \frac{V}{32.5 \hspace{0.5mm} {\rm ktons}}  \right)
\left(  \frac{d}{10 \hspace{0.5mm} {\rm kpc}}  \right)^{-2}\, , &
\label{eq:fitSKNORMAL} \\
&{\rm [DUNE-CCAre-NORMAL]}& \nonumber \\
&N_{\rm Cum} = \left( 90 \hspace{0.5mm} E_{52} + 4.5 \hspace{0.5mm} E_{52}^2 \right)
\left(  \frac{V}{40 \hspace{0.5mm} {\rm ktons}}  \right)
\left(  \frac{d}{10 \hspace{0.5mm} {\rm kpc}}  \right)^{-2}\, , &
\label{eq:fitDUNENORMAL} \\
&{\rm [JUNO-IBDp-NORMAL]}& \nonumber \\
&N_{\rm Cum} = \left( 165 \hspace{0.5mm} E_{52} + 4.5 \hspace{0.5mm} E_{52}^2 \right)
\left(  \frac{V}{20 \hspace{0.5mm} {\rm ktons}}  \right)
\left(  \frac{d}{10 \hspace{0.5mm} {\rm kpc}}  \right)^{-2}\, , &
\label{eq:fitJUNONORMAL} \\
&{\rm [IceCube-IBDp-NORMAL]}& \nonumber \\
&N_{\rm Cum} = \left( 23000 \hspace{0.5mm} E_{52} + 600 \hspace{0.5mm} E_{52}^2 \right)
\left(  \frac{V}{3.5 \hspace{0.5mm} {\rm Mtons}}  \right)
\left(  \frac{d}{10 \hspace{0.5mm} {\rm kpc}}  \right)^{-2}\, , &
\label{eq:fitIceCubeNORMAL}
\end{eqnarray}
and in the case with inverted mass hierarchy as
\begin{eqnarray}
&{\rm [SK-IBDp-InV]}& \nonumber \\
&N_{\rm Cum} = \left( 170 \hspace{0.5mm} E_{52} + 4 \hspace{0.5mm} E_{52}^2 \right)
\left(  \frac{V}{32.5 \hspace{0.5mm} {\rm ktons}}  \right)
\left(  \frac{d}{10 \hspace{0.5mm} {\rm kpc}}  \right)^{-2}\, , &
\label{eq:fitSKInV} \\
&{\rm [DUNE-CCAre-InV]}& \nonumber \\
&N_{\rm Cum} = \left( 90 \hspace{0.5mm} E_{52} + 4.5 \hspace{0.5mm} E_{52}^2 \right)
\left(  \frac{V}{40 \hspace{0.5mm} {\rm ktons}}  \right)
\left(  \frac{d}{10 \hspace{0.5mm} {\rm kpc}}  \right)^{-2}\, , &
\label{eq:fitDUNEInV} \\
&{\rm [JUNO-IBDp-InV]}& \nonumber \\
&N_{\rm Cum} = \left( 135 \hspace{0.5mm} E_{52} + 3 \hspace{0.5mm} E_{52}^2 \right)
\left(  \frac{V}{20 \hspace{0.5mm} {\rm ktons}}  \right)
\left(  \frac{d}{10 \hspace{0.5mm} {\rm kpc}}  \right)^{-2}\, , &
\label{eq:fitJUNOInV} \\
&{\rm [IceCube-IBDp-InV]}& \nonumber \\
&N_{\rm Cum} = \left( 18000 \hspace{0.5mm} E_{52} + 430 \hspace{0.5mm} E_{52}^2 \right)
\left(  \frac{V}{3.5 \hspace{0.5mm} {\rm Mtons}}  \right)
\left(  \frac{d}{10 \hspace{0.5mm} {\rm kpc}}  \right)^{-2}\, , &
\label{eq:fitIceCubeInV}
\end{eqnarray}
where $N_{\rm Cum}$, $E_{52}$, and $V$ denote the cumulative number of events, TONE in the unit of $10^{52} {\rm erg}$, and the detector volume, respectively. We note that Eqs.~\ref{eq:fitSKNORMAL} and \ref{eq:fitSKInV} with $V=220$ ktons provide the correlation for HK.

Although those formulae are useful to measure the TONE from $N_{\rm Cum}$, we need to mention two important points. First, those fits may be valid only for neutrino signals up to $\sim 1$ seconds from its first event, since our simulations do not cover the later phases. Thus, the fitting formula may need to be corrected to cover the entire phase of a CCSN. We are currently extending some of our 3D simulations to include the longer-term PNS cooling phase. Thus, we will address this issue in upcoming papers. Another important point is that the fitting formulae are directly affected by theoretical uncertainties in our CCSN models. Although our simulations provide sophisticated representations of neutrino signals from CCSN, there remain several open issues with the input physics, implying that the formulae may need to be revised. Despite these limitations, Eqs.~\ref{eq:fitSKNORMAL}~--~\ref{eq:fitIceCubeInV} should be useful for estimating the TONE in real observations.

\subsection{Retrieving neutrino spectra at CCSN sources}\label{subsec:retrieving}

In the previous section we showed that the cumulative number of events in each detector is capable of providing the TONE by means of a universal relation (correlation) which we found. As pointed out already, however, systematic errors are unavoidable with this approach due to uncertainties in the theoretical models. In addition to this, it is not easy to extract more detailed information on neutrino signals, such as the energy spectrum, which is accessible only by direct analyses with purely observed quantities.

Several methods have been proposed in the literature to reconstruct neutrino energy spectra from CCSNe. One of the common methods is by using an analytic formula, in which the spectrum is characterized by a combination of parameters. These parameters are estimated by taking statistical approaches which are very powerful for noisy data \citep[see, e.g., ][]{2008JCAP...12..006M,2017JCAP...11..036G,2018JCAP...04..040G,2018JCAP...12..006G}. On the other hand, the spectrum reconstruction with analytic formula potentially discards some important characteristics of the signal, since the analytic formula are not capable of capturing the complex features of the spectrum. It should also be pointed out that the spectrum reconstruction of all neutrino flavors is a challenging issue even for next generation detectors. Indeed, as already pointed out, detection techniques sensitive to heavy lepton neutrinos are still limited. Recently, \citet{2019PhRvD..99l3009L} proposed an interesting strategy with liquid-scintillator detectors, in which they attempt to reconstruct the energy spectra of all the flavors by combining the data from three reaction channels: IBD-p, eES (Elastic scattering on electrons), and elastic scattering on protons. However, this technique has a large systematic error in the energy $\lesssim 20$ MeV, which corresponds to the most important energy range for CCSN neutrinos. This implies that other strategies are required to reconstruct the energy spectra more accurately.

We have tackled this issue and developed a novel technique for spectrum reconstruction, in which data on multiple channels at different detectors are combined. In this section, we apply our method to neutrino events computed by SNOwGLoBES based on our 3D CCSN models and retrieve energy spectra for all neutrino flavors at the CCSN source assuming that the distance is known. Note that the Poisson noise and smearing effects due to detector responses are taken into account in this demonstration to mimic real observations. We also evaluate how accurately one can estimate the average energy and total energy of neutrinos. Although this technique can still be improved, the proposed method as it now is useful for determining how efficiently one can combine neutrino data from multiple detectors to retrieve the energy spectra of all neutrinos at the source. Below, we briefly review the essence of our method and then present the results. The details of the method are complex and should be described structurally; hence, we present them in another paper separately \citep{2020arXiv200810082N}.

\begin{figure*}
  \rotatebox{0}{
    \begin{minipage}{0.6\hsize}
        \includegraphics[width=\columnwidth]{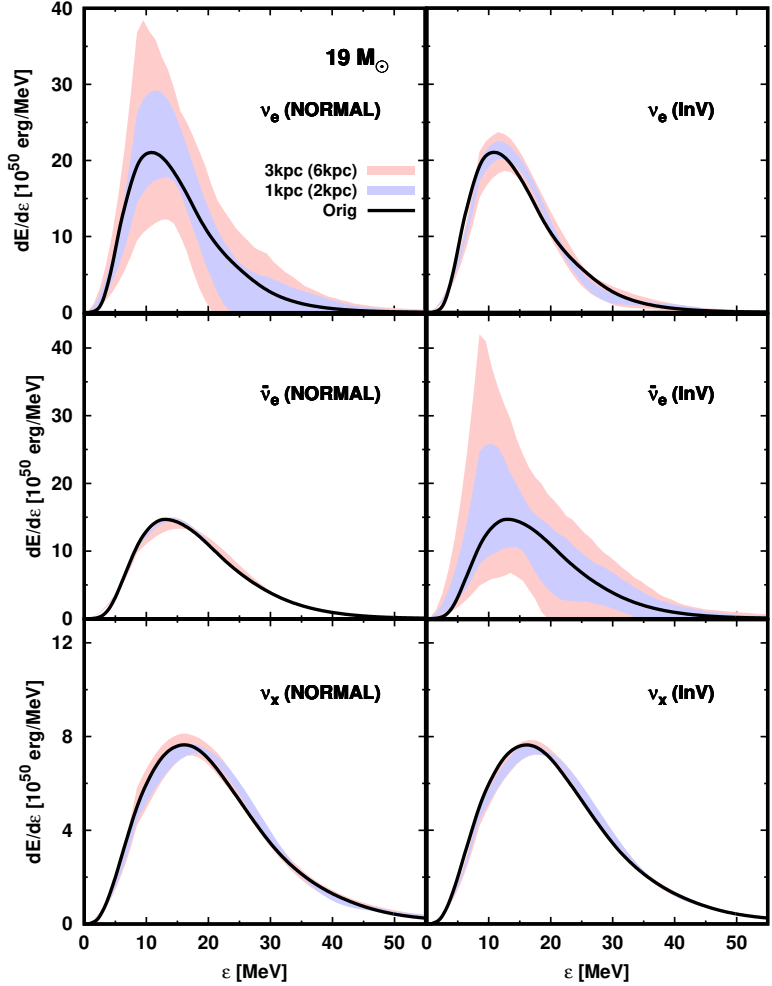}
    \caption{The time-integrated energy spectra for the 3D 19 $M_{\sun}$ model at the CCSN source retrieved by our method (see the text for more details). We focus on the data at ${\rm T} = 0.871$ s, which corresponds to the time at the end of the simulation. From top to bottom, $\nu_e$, $\bar{\nu}_e$, and $\nu_x$ neutrinos, respectively. Left and right panels show the results for the normal mass hierarchy and inverted mass hierarchy, respectively. The color distinguishes the source distance. The solid black line corresponds to the solution of the spectrum (corresponding to the result for the CCSN simulation). The shaded region corresponds to the $\sim 2 \sigma$ confidence level. The distance in parentheses denotes the corrected one including all phases of the CCSN. See the text for further details.}
    \label{graph_spectrumreconst_M19}
  \end{minipage}}
\end{figure*}

\begin{figure*}
  \rotatebox{0}{
    \begin{minipage}{0.6\hsize}
        \includegraphics[width=\columnwidth]{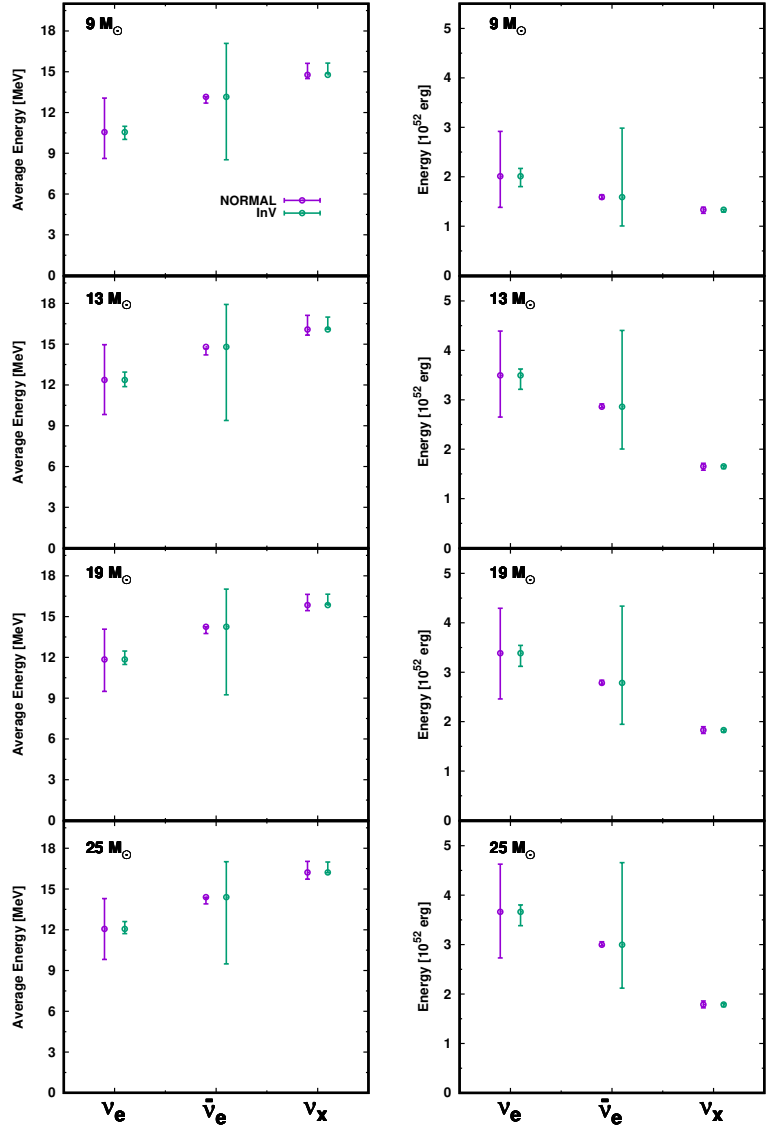}
    \caption{The average energy (left) and total energy (right) for each neutrino species retrieved by our spectrum reconstruction technique. The distance to the source is assumed to be $3 (6)$ kpc. From top to bottom, we display the results for different 3D models; 9-, 13-, 19- and 25-$M_{\sun}$. Purple and green colors denote the case with normal mass hierarchy and inverted mass hierarchy, respectively. The error bar corresponds to the $\sim 2 \sigma$ confidence level, while the open circles correspond to results from our CCSN simulations.}
    \label{graph_error_Reconst_aveE_Etot}
  \end{minipage}}
\end{figure*}

\begin{figure}
  \rotatebox{0}{
    \begin{minipage}{1.0\hsize}
        \includegraphics[width=\columnwidth]{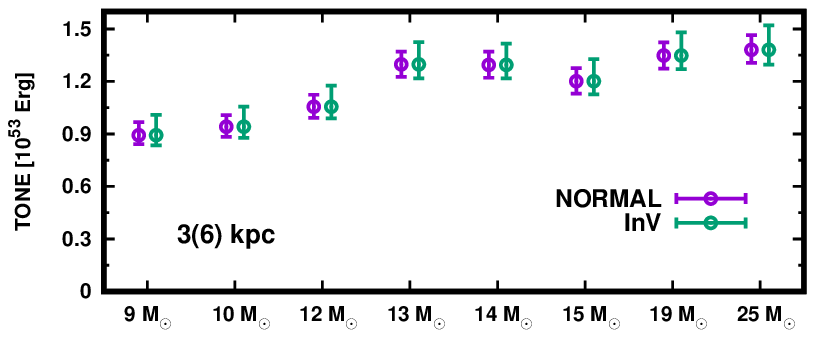}
    \caption{Total neutrino energy (TONE) estimated using energy spectra retrieved by our method. We show the results for all 3D models at the end of our simulations. Purple and green colors denote the normal mass hierarchy and inverted mass hierarchy, respectively. The error bar corresponds to the $\sim 2 \sigma$ confidence level, and the open circles corresponds to the results from CCSN simulations.}
    \label{graph_error_Reconst_fullE}
  \end{minipage}}
\end{figure}

In our method, the observed data at HK and DUNE, which have different flavor sensitivities, are adopted. Also, including HK is mandatory, unless the source is quite nearby $d \lesssim 0.5$kpc, in which case the same method can be applied using SK and DUNE. Hence, we focus only on the use of  HK and DUNE. For HK, we employ two reaction channels: IBD-p and eES. For DUNE, the CCAre channel is adopted. As we describe in Sec.~\ref{subsec:detectors}, IBD-p and CCAre are sensitive to $\bar{\nu}_e$ and $\nu_e$ neutrinos, respectively. Note that these reaction channels provide most of the events in each detector, i.e., they are high statistics. For eES, on the other hand, the reaction channel is sensitive to all neutrino flavors, although the number of counts is much smaller than via the IBD-p. It should be pointed out that we can not easily discern the flavor-dependent eES events individually, since all events are detected through a common signal (Cherenkov light from scattered electrons\footnote{Although the flavor-dependent events can not be resolved in the eES channel, the IBD-p and eES events may be distinguished by virtue of their different the angular distributions; the former is almost isotropic and the latter is forward-peaked. We also note that the addition of gadolinium greatly enhances the ability to separate these two channels. Hence, we assume in this study that we can completely distinguish the two channels. Note that we ignore the charged-current reactions with oxygen, since they are subdominant channels. However, it should be pointed out that in our method it is straightforward to add these channels.}).

As mentioned above, three reaction channels with different flavor sensitivities are employed in our approach, which is a minimum required to retrieve the neutrino spectra of all neutrino flavors at the CCSN source. 
This is due to the fact that there are three independent spectra ($\nu_e$, $\bar{\nu}_e$, and $\nu_x$) under the assumption that the energy spectra of heavy lepton neutrinos and their anti-partners are identical at the source\footnote{We note, however, that we may need to relax this assumption in real data analysis, since their neutrino-matter interactions in supernova matter are not exactly the same and the difference increase with increasing neutrino energy. As we will discuss in a forthcoming paper (Nagakura and Hotokezaka, in prep), it is possible to distinguish them, although the strategy is different from that used in this paper.}. We employ a singular value decomposition (SVD) technique \citep{1996NIMPA.372..469H}. This approach does not assume any a priori analytic formula for the spectrum shape, but evaluates the spectrum in a deterministic way. In addition, we develop an adaptive energy mesh technique with the SVD unfolding algorithm, in which the energy gridding is automatically adjusted in accordance with the energy-dependent events.

First, we apply the unfolding technique to the data for the IBD-p and CCAre, which provides us with the energy spectra of $\bar{\nu}_e$ and $\nu_e$ neutrinos in HK and DUNE, respectively. Based on these two energy spectra, we identify the eES events with $\nu_e$ and $\bar{\nu}_e$ neutrinos and then obtain the eES events of the heavy lepton neutrinos by subtracting the $\nu_e$ and $\bar{\nu}_e$ neutrino contributions from the total number of eES events\footnote{In general, the neutrino flavor state is not always the same for HK and DUNE, mainly due to the different Earth matter effects. This indicates that $\nu_e$ neutrinos at DUNE are different from those in HK (SK). Thus, one may wonder if this compromises the accuracy of the subtraction of the $\nu_e$ neutrino contribution of the eES events from the total. This is, however, not an obstacle in our method. Indeed, the Earth matter effects (and other neutrino oscillation models including non-adiabatic MSW effects and neutrino-neutrino self-interactions) can be treated self-consistently in our method. This is attributed to the fact that, again, the problem is to retrieve the three unknown energy spectra from the three independent observed data, which mathematically guarantees a unique solution. We refer readers to \citet{2020arXiv200810082N} for more details.}. It should be noted that we distinguish the spectra of $\nu_x$ and $\bar{\nu}_x$ neutrinos at the Earth, as described in Sec.~\ref{subsec:neutrinooscillations}, and doing so is mandatory in our method (since the eES cross sections are different for neutrinos and anti-neutrinos). This indicates that we need to separate the eESvents and this is possible using the assumption that $\nu_x$ and $\bar{\nu}_x$ neutrinos have identical spectra at the CCSN source. This technique requires that we iteratively search for consistent spectra using this condition. Finally, we convert the energy spectra of all neutrino flavors at the Earth to those at the CCSN source by using Eqs.~\ref{eq:flavconv_nue}--\ref{eq:flavconv_nuxb} (and the source distance).

In Fig.~\ref{graph_spectrumreconst_M19}, we show results of the spectrum reconstruction for the 3D 19 $M_{\sun}$ model. We apply our method to the energy spectrum of the cumulative number of events in each reaction channel at the end of our simulations (${\rm T} = 0.871$ s for the corresponding model). The distance to the source is set as $d=3$ kpc or $1$ kpc\footnote{Note that the SN-ratio is inversely proportional to the distance.}. It should be observed that our simulations do not cover the later PNS cooling phase (${\rm T} \gtrsim 1$ s), implying that the TONE in our CCSN models is smaller than it final value. Indeed, the TONE of all our 3D CCSN models are $\sim 10^{53}$ erg (see also Fig.~\ref{graph_error_Reconst_fullE}), which is a factor of $\sim 3-4$ smaller than the expected total emitted neutrino energy from CCSNe. This indicates that our method would be capable of providing reliable energy spectra for CCSNe with the same accuracy at roughly a factor of $\sim 2$ greater distance when all phases are included. Therefore, we also indicate the corrected distance as a reference in Fig.~\ref{graph_spectrumreconst_M19} (shown in the parentheses in the top left panel of the figure.). In this demonstration, we perform the analysis with $1000$ Poisson noise realizations for each model. The shaded region in Fig.~\ref{graph_spectrumreconst_M19} corresponds to the $\sim 2 \sigma$ confidence level.

For the normal mass hierarchy (left panels in Fig.~\ref{graph_spectrumreconst_M19}), energy spectra of $\bar{\nu}_e$ and $\nu_x$ neutrinos at the source are well retrieved. Most of the $\nu_x$ neutrinos at the source turn into $\nu_e$ neutrinos at the Earth, indicating that the spectrum reconstruction of $\nu_e$ neutrinos from CCAre events at DUNE is responsible for the $\nu_x$ neutrino spectrum at the source. By using IBD-p events in HK, we can reconstruct precisely the energy spectrum of $\bar{\nu}_e$ neutrinos at the Earth for which $\bar{\nu}_e$ and $\bar{\nu}_x$ neutrinos at the source are mixed (see Eq.~\ref{eq:flavconv_nueb}). The high precision energy spectrum of $\bar{\nu}_e$ neutrinos at the Earth (by IBD-p in HK) and that for $\bar{\nu}_x$ neutrinos at the source (using the assumption that $\bar{\nu}_x=\nu_x$ at the source, well retrieved by the CCAr data in DUNE) provides a precise $\bar{\nu}_e$ energy spectrum at the source. Incidentally, the above explanation makes clear that the IBD-p and CCAre channels in HK and DUNE, respectively, do not constrain $\nu_e$ neutrino spectra at the source; these are determined mainly through the eES channel. As shown in the figure, however, the error in the $\nu_e$ energy spectrum at the source is significantly larger than for other species. There are a few reasons for this lower precision. As already mentioned, the number of eESvents is much smaller than for IBD-p. Indeed, the SN-ratio is several times smaller than that for the IBD-p. Furthermore, we reconstruct the energy spectrum of $\nu_x$ neutrinos at the Earth by employing a subtraction process for the eES events. As a result, the $\nu_x$ signal is smaller than that for the total eES contribution. Meanwhile, the noise is determined by the total number of eES events, implying that the SN-ratio of eES events for $\nu_x$ neutrinos becomes smaller than that of the total. It should be mentioned that the cross section for the eES reactions with $\nu_e$ and $\bar{\nu}_e$ neutrinos is higher than that for heavy lepton neutrinos.  This is due to the fact that $\nu_e$ and $\bar{\nu}_e$ neutrinos interact with electrons through both charged-current and neutral-current reactions, while heavy lepton neutrinos react only through neutral currents.

In the case of the inverted mass hierarchy, on the other hand, the energy spectra of $\nu_e$ and $\nu_x$ neutrinos can be precisely retrieved. The reason is very similar to the case for the normal hierarchy. $\bar{\nu}_x$ neutrinos at the source turn into $\bar{\nu}_e$ neutrinos almost completely. This indicates that the spectral reconstruction of $\bar{\nu}_e$ neutrinos through the IBD-p channel in HK is directly connected with the precise estimation of the energy spectrum of $\bar{\nu}_x$ neutrinos at the source. This also provides the energy spectrum of $\nu_x$ neutrinos at the source, implying that the $\nu_e$ neutrino spectrum at the source can be retrieved accurately by combining the energy spectrum of the $\nu_x$ neutrinos at the source and the $\nu_e$ neutrino spectrum at the Earth, reconstructed using the CCAre channel in DUNE. Following the same argument, the energy spectrum of $\bar{\nu}_e$ neutrinos at the source reflects the large statistical error of the spectral reconstruction of $\bar{\nu}_x$ neutrinos.

In Fig.~\ref{graph_error_Reconst_aveE_Etot}, we show the results of the retrieval of the average energy and emitted total energy of each neutrino species at the source for selected models. In the analysis, the source distance is assumed at $3 (6)$ kpc and the error bar corresponds to a $\sim 2 \sigma$ confidence level. The precision of the reconstruction of the energy spectrum directly reflects the statistical uncertainty, in which the average energy and the emitted total energy of $\nu_e$ and $\bar{\nu}_e$ neutrinos are poorly retrieved for both the normal and inverted mass hierarchies, respectively. However, for other species in both oscillation models, the statistical uncertainties in both quantities are within $\sim 20 \%$ ($\sim 2 \sigma$ confidence level).

As shown above, the precision of the reconstruction of the energy spectra, average energy, and total emitted energy depends upon both the flavor and oscillation model. $\nu_x$ neutrinos at the source can be well determined in both oscillation models. This is due to the fact that either $p$ or $\bar{p}$ is close to zero for both oscillation models, implying that the reconstructed energy spectrum of $\bar{\nu}_e$ neutrinos (via IBD-p in HK) or $\nu_e$ neutrinos (via CCAre in DUNE) provides with high precision the energy spectrum of $\nu_x$ neutrinos at the source. This property is shared by other oscillation models, unless both $p$ and $\bar{p}$ are above 0.5 (which corresponds to less realistic models with less flavor mixing\footnote{There is no evidence that the Earth matter effect, non-adiabatic MSW effects, or neutrino-neutrino self-interactions, which are not taken into account in our analysis, substantially diminish the degree of flavor conversion for either neutrinos or anti-neutrinos. Hence, a small flavor conversion model seems unlikely.}). This indicates that we will probably be able to retrieve the energy spectrum of $\nu_x$ neutrinos at the source for any realistic oscillation model. This aspect is an important advantage of estimating the TONE, since the four-species-integrated energy of $\nu_x$ neutrinos is the dominant contribution. Indeed, we confirm that the error in TONE is within $\sim 20 \%$ ($\sim 2 \sigma$ confidence level) for all 3D models (see Fig.~\ref{graph_error_Reconst_fullE}) if the source is located at $3 (6)$ kpc.

Such a high precise measurement of the TONE will enable the extraction of useful physical information for CCSNe. For instance, as shown in Fig.~\ref{graph_Tevo_Cumu}, the time evolution of the cumulative number of events at each detector has a strong progenitor dependence, since it varies by a factor of a few for the different progenitors. This indicates that we will be able to place constraints on the progenitor by applying our method to the cumulative number of events in each detector up to the post-bounce time $\sim 1$ s in real observations. Indeed, as shown in our previous studies \citep{2019MNRAS.485.3153B,2020MNRAS.491.2715B}, CCSNe with lower mass progenitors tend to have lower total emitted neutrino energies, enabling the discrimination of different progenitor models using our method. It is important to remind the reader that the TONE estimated from the direct spectral reconstructions presented here can be compared with that estimated using the universal relation discussed in Sec.~\ref{subsec:correlations}.

\section{Summary and Conclusions}\label{sec:sumconc}
Despite the theoretical determination that the fluid dynamics and neutrino transport in CCSN strongly depends upon dimension, most of the previous studies of the neutrino signal have involved 1D models.  However, 3D models are clearly more realistic. This defect in previous work is due mainly to the fact that the much more computational expensive high-fidelity 3D models have only recently become available. These new 3D models exhibit a spectrum of behaviors, and most explode, indicating that 3D numerical models have made remarkable. Motivated by this development, we have performed the first systematic study of neutrino signals in terrestrial detectors for our 3D CCSN models with the aid of the SNOwGLoBES detector software.  The results reveal some distinct differences with 1D models.

In this paper, we first illuminated the differences between 1D and 3D models in the neutrino event rates in each detector (see Fig.~\ref{graph_tevoeveratecomparebet1D3D}). PNS convection is the most important reason for the differences seen. As discussed in Sec.~\ref{subsec:3DCCSNmodel} \citep[see also][]{2020MNRAS.492.5764N}, PNS convection lifts the $\nu_x$ neutrino luminosity, but decreases the average energy of the $\nu_e$ and $\bar{\nu}_e$ neutrinos compared with 1D models. The result is a systematic difference in the neutrino event rates. We found that there are interesting differences in the neutrino signals and signatures that distinguish non-exploding and exploding models, with an important role played by PNS convection. These ingredients are missing in previous studies based on 1D models.

In Sec.~\ref{subsec:Asymmetry}, we studied the angular (observer direction) dependence of the neutrino signal and quantified the angular variations. We found that the asymmetry of the angular distribution of the neutrino event rate can be quite large, reaching $\sim 100 \%$ at some time snapshots in some models (see Fig.~\ref{graph_Tevo_AsymOn}), though the angular variation in the cumulative total event number (see Fig.~\ref{graph_Tevo_AsymCum}) might be more modest. The difference in the asymmetries of the event rate and cumulative number of event is remarkable, in particular for non-exploding models (13-, 14-, and 15-$M_{\sun}$ models). This can be understood through the properties of the spiral SASI, which appears only in our non-exploding 3D models. Its spiral motions introduce distinctive temporal modulation and large angular asymmetries in the neutrino signals. On the other hand, the angular asymmetry and its directionality varies somewhat randomly with time, indicating that the time-integrated signals (i.e., the cumulative number of events) are more isotropic. It should be noted, however, that the LESA is involved in most of our 3D models and it results in coherent asymmetric neutrino emission, implying that the time integration does not substantially reduce the asymmetry. Nevertheless, we find for all our 3D models that the angular variation of the cumulative number of events is $\lesssim 20 \%$ up to the end of each simulation. It should be pointed out that the asymmetry in neutrino signals is smaller for neutrino oscillation models than when ignoring flavor conversion. This reflects a property of $\nu_x$ neutrino emissions at the source, which are emitted more isotropically than other neutrino species.

In addition, we discussed the detectability of time variations induced by the spiral SASI encoded in neutrino signals by combining results of our simulations and semi-analytic estimates (see Sec.~\ref{subsec:timevariability}). We showed that the temporal behavior can be resolved by SK, DUNE, and JUNO only if the source is located at $\lesssim 1$ kpc. The threshold distance is extended to $\sim 2$ kpc for HK. On the other hand, IceCube may be capable of resolving the temporal behavior even when $d \sim 10$ kpc, albeit with low statistics (SN-ratio is a few). We concluded that IceCube is the best detector for a temporal analysis of the spiral SASI and that the analysis of the temporal properties of the neutrino signal will yield important insights into the internal dynamics of CCSNe.

We provided Eqs.~\ref{eq:fitSKNORMAL}--\ref{eq:fitIceCubeInV} to estimate the total neutrino energy (TONE) from the cumulative number of events in each detector. This approach was designed to avail ourselves of correlations manifesting less of a progenitor dependence (see Sec.~\ref{subsec:correlations} for more details). Although there remain some uncertainties in our theoretical models, those correlations will prove useful for analyzing real observations, and would be very powerful when analyzing low-statistics neutrino signals, i.e., for distant CCSNe.  We also developed a novel method by which to retrieve energy spectra for all neutrino flavors at the CCSN source based on purely observational quantities using multiple detectors, in particular HK (SK) and DUNE (see Sec.~\ref{subsec:retrieving}). The proposed method does not a priori assume any analytic formulae. We demonstrated that the energy spectrum of $\nu_x$ neutrinos and that of either $\nu_e$ or $\bar{\nu}_e$ neutrinos at the source can be retrieved using our new method (see Fig.~\ref{graph_spectrumreconst_M19}). This indicates that the TONE can be estimated rather precisely, since the four-species-integrated energy of $\nu_x$ neutrinos is the dominant contribution to the total radiated energy. At a distance of $d=3 (6)$ kpc, we can estimate each to an accuracy of  $\sim 20 \%$ (Fig.~\ref{graph_error_Reconst_fullE}) by using this method.

Finally, we provide several caveats. The largest is that our 3D CCSN simulations should be extended to cover the late-time PNS cooling phase ($\gtrsim 1$ s), during which more than half of the total neutrino energy is emitted. Theoretical predictions of the late-time evolution are necessary to determine whether the central remnant is a neutron star or black hole \citep[for which case the signal abruptly ceases upon relativistic collapse; see, e.g.,][]{1986ApJ...300..488B,2006PhRvL..97i1101S}, and its properties.  Second, more detailed studies are called for of the sensitivities to the input physics in our CCSN models. For instance, we need to determine the nuclear equation-of-state dependence, as well as that of various neutrino-matter interactions. We leave these broad tasks to future work.

\section*{Acknowledgements}

The authors acknowledge Kate Scholberg for help in using SNOwGLoBES software.
We are also grateful for ongoing contributions to
this effort by Josh Dolence and Aaron Skinner and acknowledge Kenta Hotokezaka for profitable discussions on supernova neutrinos, Evan O'Connor regarding the equation of state, Gabriel Mart\'inez-Pinedo concerning electron capture on heavy nuclei, Tug Sukhbold and Stan Woosley for providing details concerning the initial models, and Todd Thompson regarding inelastic scattering. We acknowledge support from the U.S. Department of Energy Office of Science and the Office of Advanced Scientific Computing Research via the Scientific Discovery
through Advanced Computing (SciDAC4) program and Grant DE-SC0018297
(subaward 00009650). In addition, we gratefully acknowledge support
from the U.S. NSF under Grants AST-1714267 and PHY-1804048 (the latter
via the Max-Planck/Princeton Center (MPPC) for Plasma Physics). DR
cites partial support as a Frank and Peggy Taplin Fellow at
the Institute for Advanced Study. An award of computer time was provided 
by the INCITE program. That research used resources of the
Argonne Leadership Computing Facility, which is a DOE Office of Science 
User Facility supported under Contract DE-AC02-06CH11357. In addition, this overall research 
project is part of the Blue Waters sustained-petascale computing project,
which is supported by the National Science Foundation (awards OCI-0725070
and ACI-1238993) and the state of Illinois. Blue Waters is a joint effort
of the University of Illinois at Urbana-Champaign and its National Center
for Supercomputing Applications. This general project is also part of
the ``Three-Dimensional Simulations of Core-Collapse Supernovae" PRAC
allocation support by the National Science Foundation (under award \#OAC-1809073).
Moreover, access under the local award \#TG-AST170045
to the resource Stampede2 in the Extreme Science and Engineering Discovery
Environment (XSEDE), which is supported by National Science Foundation grant
number ACI-1548562, was crucial to the completion of this work.  Finally,
the authors employed computational resources provided by the TIGRESS high
performance computer center at Princeton University, which is jointly
supported by the Princeton Institute for Computational Science and
Engineering (PICSciE) and the Princeton University Office of Information
Technology, and acknowledge our continuing allocation at the National
Energy Research Scientific Computing Center (NERSC), which is
supported by the Office of Science of the US Department of Energy
(DOE) under contract DE-AC03-76SF00098.

\section*{DATA AVAILABILITY}
The data underlying this article will be shared on reasonable request to the corresponding author.

\appendix

\section{Energy-resolution dependence}\label{sec:resodepe} 

\begin{figure*}
  \begin{minipage}{0.8\hsize}
        \includegraphics[width=\columnwidth]{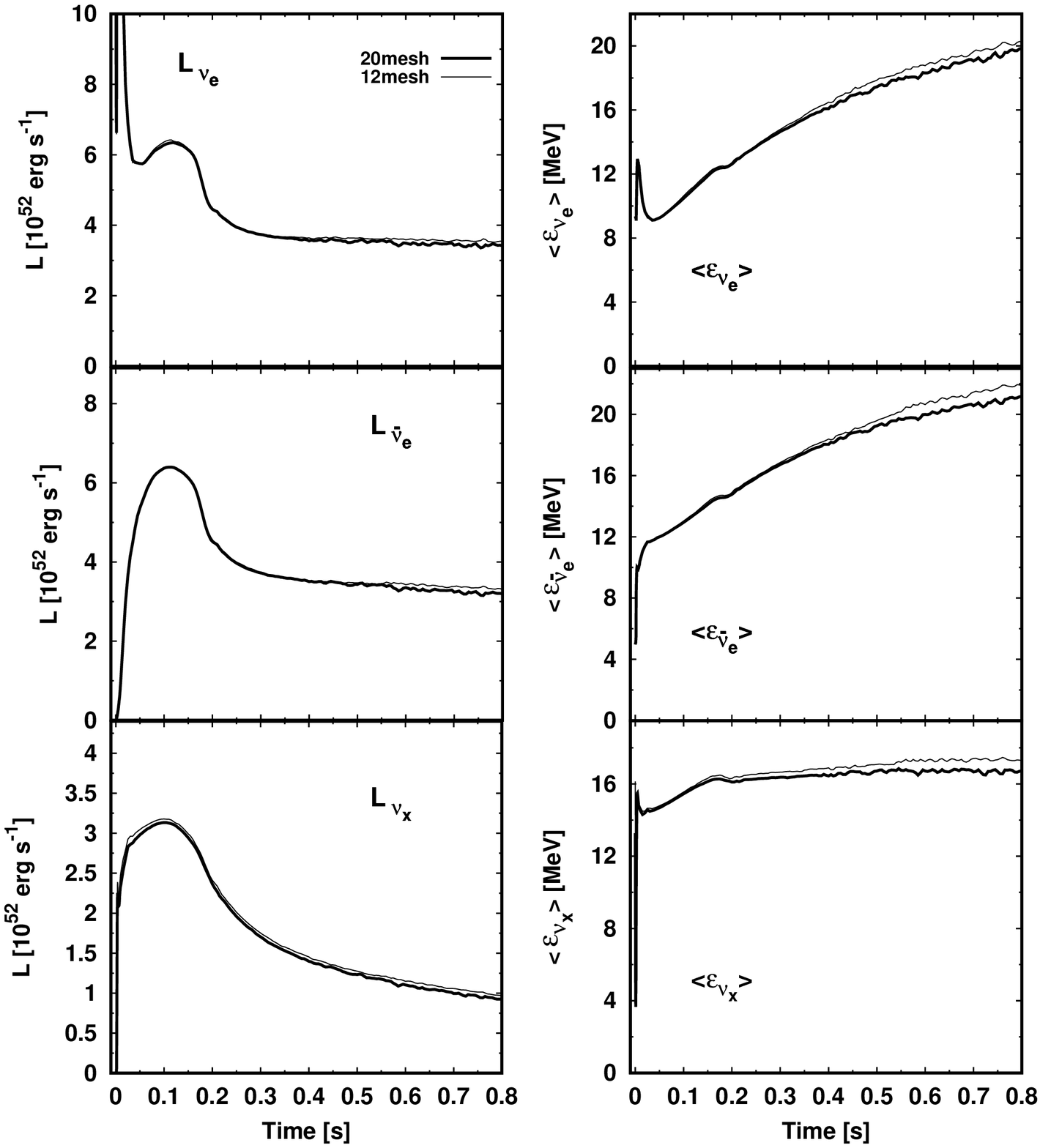}
    \caption{Same as Fig.~\ref{graph_neutrino_lumi_aveene_neutsig} but for resolution dependence in 1D 19 $M_{\sun}$ models. The thick and thin lines denote the results with 20 and 12 energy group, respectively.}
    \label{graph_neutrino_lumi_aveene_1dresodepe}
  \end{minipage}
\end{figure*}

\begin{figure}
  \rotatebox{0}{
    \begin{minipage}{0.95\hsize}
        \includegraphics[width=\columnwidth]{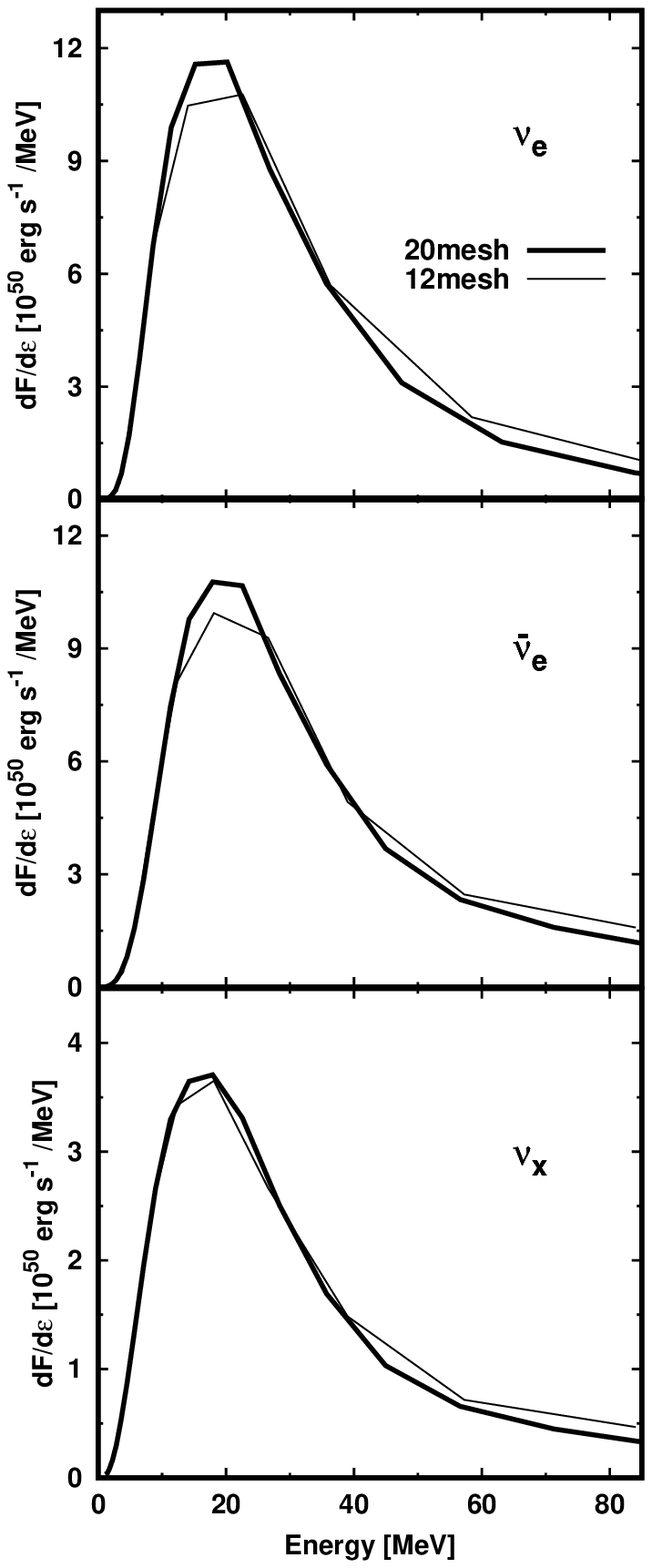}
    \caption{The energy-resolution dependence for the spectrum of neutrinos energy flux at the CCSN source for 1D 19 $M_{\sun}$ models. We compare the spectra at ${\rm T} = 600$ms. The thick and thin lines denote the resutls with 20 and 12 energy group, respectively.}
    \label{graph_eneSpect_1Dresodepe}
  \end{minipage}}
\end{figure}

\begin{figure}
  \rotatebox{0}{
    \begin{minipage}{1.0\hsize}
        \includegraphics[width=\columnwidth]{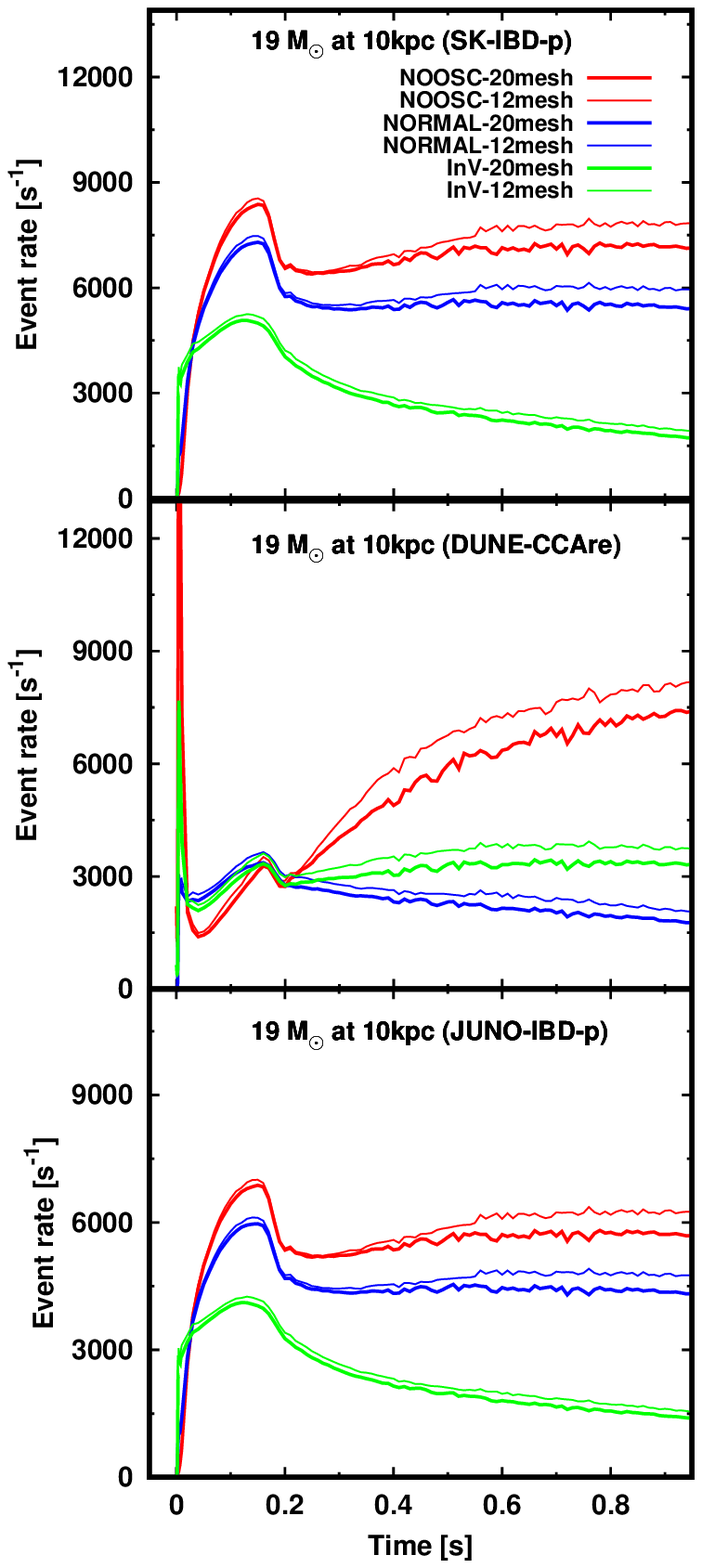}
    \caption{Same as Fig.~\ref{graph_tevoeveratecomparebet1D3D} but for resolution dependence in 1D 19 $M_{\sun}$ model. The color distinguishes neutrino oscillation models. The thick and thin lines denote the results with 20 and 12 energy group, respectively.}
    \label{graph_1Dresodepe_deterate}
  \end{minipage}}
	\end{figure}

In this appendix, we address the energy-resolution dependence of our CCSN models by employing the 1D 19 $M_{\sun}$ progenitor. We ran two different simulations of 20 and 12 energy groups. This study helps us understand how the energy resolution in our 3D CCSN simulations (with 12 energy groups) affects the outcome.

In Fig.~\ref{graph_neutrino_lumi_aveene_1dresodepe}, we show the time evolution of neutrino luminosity and average energy. As shown in this figure, the resolution dependence is weak (a few percents); those energy-integrated quantities are not sensitive to the energy resolution. 
In Fig.~\ref{graph_eneSpect_1Dresodepe}, we compare the energy flux spectra of neutrinos at the CCSN source. We find that the spectrum at peak ($\sim 10$ MeV) of the 12-energy group model is slightly broader than that of the 20-energy group model and that there is a slight excess
iof the 12-group model at the high energy tail in the spectrum.

Although the excess of the high-energy tail looks minor in Fig.~\ref{graph_eneSpect_1Dresodepe}, this affects the detection rate, which is shown in Fig.~\ref{graph_1Dresodepe_deterate}. The detection rate in the 12-energy group run is systematically higher than that of 20-energy group run. This is mainly because the cross section of each reaction is an increasing function with energy. It should be noted, however, that the discrepancy is $\lesssim 10 \%$ level, and the high-energy component in the 3D model is smaller than that of 1D (see in Sec.~\ref{subsec:ProgenitorDependence} for more details), implying that the resolution dependence in 3D would be smaller than in 1D.

%\clearpage

%%%%%%%%%%%%%%%%%%%%%%%%%%%%%%%%%%%

%%%%%%%%%%%%%%%%%%%% REFERENCES %%%%%%%%%%%%%%%%%%

% The best way to enter references is to use BibTeX:

\bibliographystyle{mnras}
\bibliography{bibfile}

%%%%%%%%%%%%%%%%%%%%%%%%%%%%%%%%%%%%%%%%%%%%%%%%%%

%%%%%%%%%%%%%%%%% APPENDICES %%%%%%%%%%%%%%%%%%%%%

%\appendix

%\section{Some extra material}

%If you want to present additional material which would interrupt the flow of the main paper,
%it can be placed in an Appendix which appears after the list of references.

%%%%%%%%%%%%%%%%%%%%%%%%%%%%%%%%%%%%%%%%%%%%%%%%%%

% Don't change these lines
\bsp	% typesetting comment
\label{lastpage}
\end{document}